\documentclass[12pt]{article}
\usepackage{xr-hyper}
\usepackage{changepage}

    \usepackage[dvipsnames]{xcolor}
    \usepackage[T1]{fontenc}
	\usepackage[utf8]{inputenc}
	\usepackage[english]{babel}
	\usepackage{amsfonts}
	\usepackage{amssymb}
	    \usepackage{amsmath}
	\usepackage{adjustbox}
	\usepackage{threeparttable}
\usepackage{booktabs}	
\usepackage[normalem]{ulem}
	\usepackage{enumerate}
	\usepackage{bbm}
	\usepackage{mathtools}
	\usepackage{subfig}
	\usepackage{booktabs}	
	\usepackage{setspace}
	\usepackage{graphicx}
	\usepackage{fullpage}
	
	\usepackage{soul}
	\usepackage{xspace}
	\usepackage{csquotes}
	\usepackage[margin = 1in]{geometry}
	\usepackage{placeins}
	\usepackage[footfont=normalsize,font=normalsize]{floatrow}
	\usepackage{lscape}
	\usepackage{longtable}
	\usepackage{mathabx}
    \usepackage{accents}
    \usepackage{float}

    \usepackage{tikz}
    \usetikzlibrary{calc,intersections}

	\usepackage[longnamesfirst]{natbib}
	\usepackage{bibunits}
	\defaultbibliography{Bibliography.bib}  
	\defaultbibliographystyle{aer}

	\usepackage{hyperref}
	\hypersetup{
    colorlinks=true,
    linkcolor=Red,
    filecolor=magenta,      
    urlcolor=blue,
    citecolor = Blue
}
	\usepackage{cleveref}

    \usepackage{rotating}
	\usepackage[framemethod=tikz]{mdframed}

    \usepackage{tcolorbox}
    \newtcbox{\feedback}{nobeforeafter,colframe=black,colback=white,boxrule=0.5pt,arc=2pt,
      boxsep=0pt,left=2pt,right=2pt,top=2pt,bottom=2pt,tcbox raise base}
      
    \usepackage{amsthm}

    \newtheorem{asm}{Assumption}

    \theoremstyle{definition}

\usepackage{array}
\newcolumntype{L}[1]{>{\raggedright\let\newline\\\arraybackslash}m{#1}}
\newcolumntype{C}[1]{>{\centering\let\newline\\\arraybackslash\hspace{0pt}}m{#1}}
\newcolumntype{R}[1]{>{\raggedleft\let\newline\\\arraybackslash\hspace{0pt}}m{#1}}

\usepackage{ragged2e}
\newlength\ubwidth

\newcommand\numberthis{\addtocounter{equation}{1}\tag{\theequation}}

	\newcommand{\bracks}[1]{\left[#1\right]}
	\newcommand{\expe}[1]{\mathbb{E}\bracks{#1}}

	\newcommand{\parens}[1]{\left(#1\right)}

	\newcommand{\betahat}{\ensuremath{\widehat{\beta}}}

\newcommand{\normnot}[2]{\mathcal{N}\parens{#1,\,#2}}

\newcommand{\betahatpost}{\betahat_{post}}

\newcommand{\betahatpre}{\betahat_{pre}}

\newcommand{\betapost}{\beta_{post}}

\title{What's Trending in Difference-in-Differences? \\ A Synthesis of the Recent Econometrics Literature\thanks{We thank Brant Callaway, Bruno Ferman, Andreas Hagemann, Kevin Lang, David McKenzie, and David Sch\"{o}nholzer for helpful comments, and Scott Barkowski for suggesting the title.}}

\author{Jonathan Roth\thanks{Brown University. \href{mailto:jonathanroth@brown.edu}{jonathanroth@brown.edu}} \and Pedro H. C. Sant'Anna\thanks{Microsoft and Vanderbilt University. \href{mailto:pedro.h.santanna@vanderbilt.edu}{pedro.h.santanna@vanderbilt.edu}} \and Alyssa Bilinski\thanks{Brown University. \href{mailto:alyssa\_bilinski@brown.edu}{alyssa\_bilinski@brown.edu}} \and John Poe\thanks{University of Michigan. \href{mailto:john@johndavidpoe.com}{john@johndavidpoe.com}}}

\begin{document}
\maketitle
\begin{abstract}
    This paper synthesizes recent advances in the econometrics of difference-in-differences (DiD) and provides concrete recommendations for practitioners. We begin by articulating a simple set of ``canonical'' assumptions under which the econometrics of DiD are well-understood. We then argue that recent advances in DiD methods can be broadly classified as relaxing some components of the canonical DiD setup, with a focus on $(i)$ multiple periods and variation in treatment timing, $(ii)$ potential violations of parallel trends, or $(iii)$ alternative frameworks for inference. Our discussion highlights the different ways that the DiD literature has advanced beyond the canonical model, and helps to clarify when each of the papers will be relevant for empirical work. We conclude by discussing some promising areas for future research.
\end{abstract}



\section{Introduction}

Differences-in-differences (DiD) is one of the most popular methods in the social sciences for estimating causal effects in non-experimental settings. The last few years have seen a dizzying array of new methodological papers on DiD and related designs, making it challenging for practitioners to keep up with rapidly evolving best practices. Furthermore, the recent literature has addressed a variety of different components of DiD analyses, which has made it difficult even for experts in the field to understand how all of the new developments fit together. In this paper, we attempt to synthesize some of the recent advances on DiD and related designs and to provide concrete recommendations for practitioners. 

Our starting point in Section \ref{sec: basic model} is the ``canonical'' difference-in-differences model, where two time periods are available, there is a treated population of units that receives a treatment of interest beginning in the second period, and a comparison population that does not receive the treatment in either period. The key identifying assumption is that the average outcome among the treated and comparison populations would have followed ``parallel trends'' in the absence of treatment. We also assume that the treatment has no causal effect before its implementation (no anticipation). Together, these assumptions allow us to identify the average treatment effect on the treated (ATT). If we observe a large number of independent clusters from the treated and comparison populations, the ATT can be consistently estimated using a two-way fixed effects (TWFE) regression specification, and clustered standard errors provide asymptotically valid inference. 

In practice, DiD applications typically do not meet all of the requirements of the canonical DiD setup. The recent wave of DiD papers have each typically focused on relaxing one or two of the key assumptions in the canonical framework while preserving the others. We taxonomize the recent DiD literature by characterizing which of the key assumptions in the canonical model are relaxed. We focus on recent advances that $(i)$ allow for multiple periods and variation in treatment timing (Section \ref{sec:timing}); $(ii)$ consider potential violations of parallel trends (Section \ref{sec: pt violated}); or $(iii)$ depart from the assumption of observing a sample of many independent clusters sampled from a super-population (Section \ref{sec: inference}). Section \ref{sec: other topics} briefly summarizes some other areas of innovation. In the remainder of the Introduction, we briefly describe each of these strands of literature.

\textbf{Multiple periods and variation in treatment timing:} One strand of the DiD literature has focused on settings where there are more than two time periods and units are treated at different point in times. Multiple authors have noted that the coefficients from standard TWFE models may not represent a straightforward weighted average of unit-level treatment effects when treatment effects are allowed to be heterogeneous. In short, TWFE regressions make both ``clean'' comparisons between treated and not-yet-treated units as well as ``forbidden'' comparisons between units who are both already-treated. When treatment effects are heterogeneous, these ``forbidden'' comparisons potentially lead to severe drawbacks such as TWFE coefficients having the opposite sign of all individual-level treatment effects due to ``negative weighting'' problems. Even if all of the weights are positive, the weights ``chosen'' by TWFE regressions may not correspond with the most policy-relevant parameter. 

We discuss a variety of straightforward-to-implement strategies that have been proposed to bypass the limitations associated with TWFE regressions and estimate causal parameters of interest under rich sources of treatment effect heterogeneity. These procedures rely on generalizations of the parallel trends assumption to the multi-period setting. A common theme is that these new estimators isolate ``clean'' comparisons between treated and not-yet-treated groups, and then aggregate them using user-specified weights to estimate a target parameter of economic interest. We discuss differences between some of the recent proposals --- such as the exact comparison group used and the generalization of the parallel trends assumption needed for validity --- and provide concrete recommendations for practitioners. We also briefly discuss extensions to more complicated settings such as when treatments turn on-and-off over time or are non-binary. 


\textbf{Non-parallel trends:} A second strand of the DiD literature focuses on the possibility that the parallel trends assumption may be violated. One set of papers considers the setting where parallel trends holds only conditional on observed covariates, and proposes new estimators that are valid under a conditional parallel trends assumption. However, even if one conditions on observable covariates, there are often concerns that the necessary parallel trends assumption may be violated due to time-varying unobserved confounding factors. It is therefore common practice to test for pre-treatment differences in trends (``pre-trends'') as a test of the plausibility of the (conditional) parallel trends assumption. While intuitive, researchers have identified at least three issues with this pre-testing approach. First, the absence of a significant pre-trend does not necessarily imply that parallel trends holds; in fact, these tests often have low power. Second, conditioning the analysis on the result of a pre-test can introduce additional statistical distortions from a selection effect known as pre-test bias. Third, if a significant difference in trends is detected, we may still wish to learn something about the treatment effect of interest. 

Several recent papers have therefore suggested alternative methods for settings where there is concern that parallel trends may be violated. One class of solutions involves modifications to the common practice of pre-trends testing to ensure that the power of pre-tests is high against relevant violations of parallel trends. A second class of solutions has proposed methods that remain valid under certain types of violations of parallel trends, such as when the post-treatment violation of parallel trends is assumed to be no larger than the maximal pre-treatment violation of parallel trends, or when there are non-treated groups that are known to be more/less affected by the confounds as the treated group. These approaches allow for a variety of robustness and sensitivity analyses which are useful in a wide range of empirical settings, and we discuss them in detail. 


\textbf{Alternative sampling assumptions:} A third strand of the DiD literature discusses alternatives to the classical ``sampling-based'' approach to inference with a large number of clusters. One topic of interest is inference procedures in settings with a small number of treated clusters. Standard cluster-robust methods assume that there is a large number of both treated and untreated clusters, and thus can perform poorly in this case. A variety of alternatives with better properties have been proposed for this case, including permutation and bootstrap procedures. These methods typically either model the dependence of errors across clusters, or alternatively place restrictions on the treatment assignment mechanism. We briefly highlight these approaches and discuss the different assumptions needed for them to perform well.

Another direction that has been explored relates to conducting ``design-based'' inference for DiD. Canonical approaches to inference suppose that we have access to a sample of independently-drawn clusters from an infinite super-population. However, it is not always clear how to define the super-population, or to determine the appropriate level of clustering. Design-based approaches address these issues by instead treating the source of randomness in the data as coming from the stochastic assignment of treatment, rather than sampling from an infinite super-population. Although design-based approaches have typically been employed in the case of randomized experiments, recent work has extended this to the case of ``quasi-experimental'' strategies like DiD. Luckily, the message of this literature is positive, in the sense that methods that are valid from the canonical sampling-based view are typically also valid from the design-based view as well. The design-based approach also yields the clear recommendation that it is appropriate to cluster standard errors at the level at which treatment is independently assigned.

\paragraph{Other topics:} We conclude by briefly touching on some other areas of focus within the DiD literature, as well as highlighting some areas for future research. Examples include using DiD to estimate distributional treatment effects; settings with quasi-random treatment timing; spillover effects; estimating heterogeneous treatment effects; and connections between DiD and other panel data methods. \\

Overall, the growing DiD econometrics literature emphasizes the importance of clarity and precision in a researcher's discussion of his or her assumptions, comparison group and time frame selection, causal estimands, estimation methods, and robustness checks.  When used in combination with context-specific information, these new methods can both improve the validity and interpretability of DiD results and more clearly delineate their limitations.

Given the vast literature on DiD, our goal is not to be comprehensive, but rather to give a clean presentation of some of the most important directions the literature has gone. Wherever possible, we try to give clear practical guidance for applied researchers, concluding each section with practical recommendations for applied researchers. For reference, we include Table \ref{tbl: checklist}, which contains a checklist for a practitioner implementing a DiD analysis, and Table \ref{tbl:packages}, which lists R and Stata packages for implementing many of the methods described in this paper. 

\section{The Basic Model\label{sec: basic model}}

This section describes a simple two-period setting in which the econometrics of DiD are well-understood. Although this ``canonical'' setting is arguably too simple for most applications, clearly articulating the assumptions in this setup serves as a useful baseline for understanding recent innovations in the DiD literature.

\subsection{Treatment Assignment and Timing\label{subsec:treatment assignment}}

Consider a model in which there are two time periods, $t= 1,2$. Units indexed by $i$ are drawn from one of two populations. Units from the treated population $(D_i =1)$ receive a treatment of interest between period $t=1$ and $t=2$, whereas units from the untreated (a.k.a. comparison or control) population $(D_i=0)$ remain untreated in both time periods. The econometrician observes an outcome $Y_{i,t}$ and treatment status $D_i$ for a panel of units, $i= 1,...,N$ and $t = 1,2$. For example $Y_{i,t}$ could be the fraction of people with insurance coverage in state $i$ in year $t$, while $D_i$ could be an indicator for whether the state expanded Medicaid in year 2. Although DiD methods also accommodate the case where only repeated cross-sectional data is available, or where the panel is unbalanced, we focus on the simpler setup with balanced panel data for ease of exposition.

\subsection{Potential Outcomes and Target Parameter}\label{subsec:Potential outcome basic}

We adopt a potential outcomes framework for the observed outcome, as in, e.g., \citet{Rubin1974} and \citet{Robins1986}. Let $Y_{i,t}(0,0)$ denote unit $i$'s potential outcome in period $t$ if $i$ remains untreated in both periods. Likewise, let $Y_{i,t}(0,1)$ denotes unit $i$'s potential outcome in period $t$ if $i$ is untreated in the first period but exposed to treatment by the second period. To simplify notation we will write $Y_{i,t}(0) = Y_{i,t}(0,0)$ and $Y_{i,t}(1) = Y_{i,t}(0,1)$, but it will be useful for our later discussion to make clear that these potential outcomes in fact correspond with a path of treatments. As is usually the case, due to the fundamental problem of causal inference \citep{holland_statistics_1986}, we only observe one of the two potential outcomes for each unit $i$. That is, the observed outcome is given by $Y_{i,t} = D_i Y_{i,t}(1) + (1-D_i) Y_{i,t}(0)$. This potential outcomes framework implicitly encodes the stable unit treatment value assumption (SUTVA) that unit $i$'s outcomes do not depend on the treatment status of unit $j\neq i$, which rules out spillover and general equilibrium effects.

The causal estimand of primary interest in the canonical DiD setup is the average treatment effect on the treated (ATT) in period $t=2$,
$$ \tau_{2} = \expe{ Y_{i,2}(1) - Y_{i,2}(0) \,|\, D_i = 1 }.$$
It simply measures the average causal effect on treated units in the period that they are treated ($t=2$). In our motivating example, $\tau_2$ would be the average effect of Medicaid expansion on insurance coverage in period 2 for the states who expanded Medicaid. 

\subsection{The Parallel Trends Assumption and Identification}

The challenge in identifying $\tau_2$ is that the untreated potential outcomes, $Y_{i,2}(0)$, are never observed for the treated group ($D_i=1$). Difference-in-differences methods overcome this identification challenge via assumptions that allow us to impute the mean counterfactual untreated outcomes for the treated group by using (a) the change in outcomes for the untreated group and (b) the baseline outcomes for the treated group. The key assumption for identifying $\tau_{2}$ is the parallel trends assumption, which intuitively states that the average outcome for the treated and untreated populations would have evolved in parallel if treatment had not occurred.

\begin{asm}[Parallel Trends] \label{asm: parallel-trends-2-periods}
\begin{equation}
\expe{ Y_{i,2}(0) - Y_{i,1}(0) \,|\, D_i = 1 } = \expe{ Y_{i,2}(0) - Y_{i,1}(0) \,|\, D_i = 0 } . \label{eqn: pt - 2 periods}
\end{equation}

\end{asm}

\noindent In our motivating example, the parallel trends assumption says that the average change in insurance coverage for expansion and non-expansion states would have been the same in the absence of the Medicaid expansion.

The parallel trends assumption can be rationalized by imposing a particular generative model for the untreated potential outcomes. If $Y_{i,t}(0) = \alpha_i + \phi_t + \epsilon_{i,t}$, where $\epsilon_{i,t}$ is mean-independent of $D_i$, then Assumption \ref{asm: parallel-trends-2-periods} holds. Note that this model allows treatment to be assigned non-randomly based on characteristics that affect the level of the outcome $(\alpha_i)$, but requires the treatment assignment to be mean-independent of variables that affect the  \textit{trend} in the outcome ($\epsilon_{i,t}$). In other words, parallel trends allows for the presence of selection bias, but the bias from selecting into treatment must be the same in period $t=1$ as it is in period $t=2$.

Another important and often hidden assumption required for identification of $\tau_2$ is the no-anticipation assumption, which states that the treatment has no causal effect prior to its implementation. This is important for identification of $\tau_2$, since otherwise the changes in the outcome for the treated group between period 1 and 2 could reflect not just the causal effect in period $t=2$ but also the anticipatory effect in period $t=1$ \citep{Abbring2003, malani_interpreting_2015}.

\begin{asm}[No anticipatory effects] \label{asm: no anticipation}
$Y_{i,1}(0) = Y_{i,1}(1)$ for all $i$ with $D_i =1$.
\end{asm}
\noindent In our ongoing example, this implies that in years prior to Medicaid expansion, insurance coverage in states that expanded Medicaid was not affected by the upcoming Medicaid expansion.

Under the parallel trends and no-anticipation assumptions, the ATT in period 2 ($\tau_2$) is identified. To see why this is the case, observe that by re-arranging terms in the parallel trends assumption (see equation (\ref{eqn: pt - 2 periods})), we obtain 
$$\expe{ Y_{i,2}(0) \,|\, D_i = 1} =  \expe{Y_{i,1}(0) \,|\, D_i = 1 } + \expe{ Y_{i,2}(0) - Y_{i,1}(0) \,|\, D_i = 0 }.$$
\noindent Further, by the no anticipation assumption,  $ \expe{Y_{i,1}(0) \,|\, D_i = 1 } = \expe{Y_{i,1}(1) \,|\, D_i = 1}$.\footnote{For the identification argument, it suffices to impose only that $ \expe{Y_{i,1}(0) \,|\, D_i = 1 } = \expe{Y_{i,1}(1) \,|\, D_i = 1}$ directly, i.e. that there is no anticipation on average, which is slightly weaker than Assumption \ref{asm: no anticipation}. We focus on Assumption \ref{asm: no anticipation} for ease of exposition (especially when we extend it to the staggered case below). \label{fn: no anticipation assumption}} It follows that 
\begin{align*}
\expe{ Y_{i,2}(0) \,|\, D_i = 1} &=  \expe{Y_{i,1}(1) \,|\, D_i = 1 } + \expe{ Y_{i,2}(0) - Y_{i,1}(0) \,|\, D_i = 0 } \\
&=  \expe{Y_{i,1} \,|\, D_i = 1 } + \expe{ Y_{i,2} - Y_{i,1} \,|\, D_i = 0 },
\end{align*}

\noindent where the second equality uses the fact that we observe $Y(1)$ for treated units and $Y(0)$ for untreated units. The previous display shows that we can infer the \textit{counterfactual} average outcome for the treated group by taking its observed pre-treatment mean and adding the change in mean for the untreated group. Since we observe $Y(1)$ for the treated group directly, it follows that $\tau_2 = \expe{Y_{i,2}(1) - Y_{i,2}(0) \,|\, D_i = 1}$ is identified as
\begin{equation}
\tau_2 = \underbrace{\expe{Y_{i,2} - Y_{i,1} \,|\,D_i = 1}}_{\text{ Change for $D_i=1$}} - \underbrace{\expe{Y_{i,2} - Y_{i,1} \,|\, D_i = 0}}_{\text{Change for $D_i = 0$}},  \label{eqn: identification-tau2-2-periods}  
\end{equation}
\noindent i.e. the ``difference-in-differences'' of population means!

\subsection{Estimation and Inference}\label{subse: DiD estimation and inference}

Equation (\ref{eqn: identification-tau2-2-periods}) gives an expression for $\tau_2$ in terms of a ``difference-in-differences'' of population expectations. Therefore, a natural way to estimate $\tau_2$ is to replace expectations with their sample analogs,
$$\widehat{\tau}_2 = (\overline{Y}_{t=2,D=1} - \overline{Y}_{t=1,D=1}) - (\overline{Y}_{t=2,D=0} - \overline{Y}_{t=1,D=0}),$$

\noindent where $\overline{Y}_{t=t',D=d}$ is the sample mean of $Y$ for treatment group $d$ in period $t'$.

Although these sample means could be computed ``by hand'', an analogous way of computing $\widehat{\tau}_2$, which facilitates the computation of standard errors, is to use the two-way fixed effects (TWFE) regression specification 

\begin{equation}
Y_{i,t} = \alpha_i + \phi_t + (1[t=2] \cdot D_i) \beta  + \epsilon_{i,t}, \label{eqn: TWFE-2-periods}
\end{equation}

\noindent which regresses the outcome $Y_{i,t}$ on an individual fixed effect, a time fixed effect, and an interaction of a post-treatment indicator with treatment status.\footnote{\label{fn: group_FE} With a balanced panel, the OLS coefficient on $\beta$ is also numerically identical to the coefficient from a regression that replaces the fixes effects with a constant, a treatment indicator, a second-period indicator, and the treatment $\times$ second-period interaction,
$$Y_{i,t} = \alpha + D_i \theta + 1[t=2] \zeta + (1[t=2] \cdot D_i) \beta + \varepsilon_{i,t}.$$
\noindent The latter regression generalizes to repeated cross-sectional data.
} In this canonical DiD setup, it is straightforward to show that the ordinary least squares (OLS) coefficient $\betahat$ is equivalent to $\widehat{\tau}_2$.

OLS estimates of $\betahat$ from (\ref{eqn: TWFE-2-periods}) provide consistent estimates and asymptotically valid confidence intervals of $\tau_2$ when Assumptions \ref{asm: parallel-trends-2-periods} and \ref{asm: no anticipation} are combined with the assumption of independent sampling.

\begin{asm}\label{asm: random sampling - two periods}
Let $W_i = ( Y_{i,2}, Y_{i,1}, D_i)'$ denote the vector of outcomes and treatment status for unit $i$. We observe a sample of $N$ i.i.d. draws $W_i \sim F$ for some distribution $F$ satisfying parallel trends.
\end{asm}

Under Assumptions \ref{asm: parallel-trends-2-periods}-\ref{asm: random sampling - two periods} and mild regularity conditions, 

$$ \sqrt{n}(\betahat - \tau_2) \rightarrow_d \normnot{0}{\sigma^2},$$

\noindent in the asymptotic as $N\rightarrow\infty$ and $T$ is fixed. The variance $\sigma^2$ is consistently estimable using standard clustering methods that allow for arbitrary serial correlation at the unit level \citep{liang_longitudinal_1986, arellano_practitioners_1987, Wooldridge2003,bertrand_how_2004}. The same logic easily extends to cases where the observations are individual units who are members of independently-sampled clusters (e.g. states), and the standard errors are clustered at the appropriate level, provided that the number of treated and untreated clusters both grow large. Constructing consistent point estimates and asymptotically valid confidence intervals is thus straightforward via OLS.

Having introduced all of the components of the ``canonical'' DiD model, we now discuss the ways that different strands of the recent DiD have relaxed each of these components.

\section{Relaxing assumptions on treatment assignment and timing}\label{sec:timing}

Several recent papers have focused primarily on relaxing the baseline assumptions about treatment assignment and timing discussed in Section \ref{sec: basic model}. A topic of considerable attention has been settings where there are more than two periods, and units adopt a treatment of interest at different points in time. For example, in practice different states expanded Medicaid in different years. We provide an overview of some of the key developments in the literature, and refer the reader to the review by \citet{de_chaisemartin_two-way-survey_2021} for additional details.  

\subsection{Generalized model with staggered treatment adoption}

Several recent papers have focused on relaxing the timing assumptions discussed in Section \ref{sec: basic model}, while preserving the remaining structure of the stylized model (i.e., parallel trends, no anticipation, and independent sampling). Since most of the recent literature considers a setup in which treatment is an absorbing state, we start with that framework; in Section \ref{subsec: related timing extensions}, we discuss extensions to the case where treatment can turn on and off or is non-binary. We introduce the following assumptions and notation, which captures the primary setting studied in this literature. 

\paragraph{Treatment timing.} There are $T$ periods indexed by $t = 1,...,T$, and units can receive a binary treatment of interest in any of the periods $t>1$. Treatment is an absorbing state, so that once a unit is treated they remain treated for the remainder of the panel. We denote by $D_{i,t}$ an indicator for whether unit $i$ receives treatment in period $t$, and $G_i = \min \{ t: D_{i,t} =1\}$ the earliest period at which unit $i$ has received treatment. If $i$ is never treated during the sample, then $G_i = \infty$. Treatment is an absorbing state, so that $D_{i,t} = 1$ for all $t \geq G_i$. Thus, for example, a state that first expanded Medicaid in 2014 would have $G_i = 2014$, a state that first expanded Medicaid in 2015 would have $G_i = 2015$, and a state that has not expanded Medicaid by time $t=T$ would have $G_i = \infty$.  

\paragraph{Potential outcomes.} We extend the potential outcomes framework introduced above to the multi-period setting. Let $\textbf{0}_s$ and $\textbf{1}_s$ denote $s$-dimensional vectors of zeros and ones, respectively. We denote unit $i$'s potential outcome in period $t$ if they were first treated at time $g$ by $Y_{i,t}(\textbf{0}_{g-1}, \textbf{1}_{T-g+1})$, and denote by $Y_{i,t}(\textbf{0}_{T})$ their ``never-treated'' potential outcome. This notation again makes explicit that potential outcomes can depend on the entire \textit{path} of treatment assignments. Since we have assumed that treatment ``stays on'' once it is turned on, the entire path of potential outcomes is summarized by the first treatment date $(g)$, and so to simplify notation we can index potential outcomes by treatment starting time: $Y_{i,t}(g) = Y_{i,t}(\textbf{0}_{g-1}, \textbf{1}_{T-g+1})$ and $Y_{i,t}(\infty) = Y_{i,t}(\textbf{0}_{T})$.\footnote{If one were to map this staggered potential outcome notation to the one used in canonical DiD setups, we would write $Y_{i,t}(2)$ and $Y_{i,t}(\infty)$ instead of $Y_{i,t}(1)$ and $Y_{i,t}(0)$ as defined in Section \ref{subsec:Potential outcome basic}, respectively. We use the $Y_{i,t}(0),Y_{i,t}(1)$ notation in Section \ref{subsec:Potential outcome basic} because it is likely more familiar to the reader, and widely used in the literature on the canonical model.} Thus, for example, $Y_{i,2016}(2014)$ would represent the insurance coverage in state $i$ in 2016 if they had first expanded Medicaid in 2014.

\paragraph{Parallel trends.} 

There are several ways to extend the canonical parallel trends assumption to the staggered setting. The simplest extension of the parallel trends assumption to the staggered case requires that the two-group, two-period version of parallel trends holds for all combinations of periods and all combinations of ``groups'' treated at different times.

\begin{asm}[Parallel trends for staggered setting] \label{asm: pt in staggered case - all groups}
For all $t \neq t'$ and $g \neq g'$, 
\begin{equation}
\expe{ Y_{i,t}(\infty) - Y_{i,t'}(\infty) \,|\, G_i = g } =  \expe{ Y_{i,t}(\infty) - Y_{i,t'}(\infty) \,|\, G_i = g' }. \label{eqn: pt for groups} 
\end{equation} 
\end{asm}
\noindent This assumption imposes that in the counterfactual where treatment had not occurred, the average outcomes for all adoption groups would have evolved in parallel. Thus, for example, Assumption \ref{asm: pt in staggered case - all groups} would imply that --- if there had been no Medicaid expansions --- insurance rates would have evolved in parallel on average for all groups of states that adopted Medicaid expansion in different years, including those who never expanded Medicaid.

Several variants of Assumption \ref{asm: pt in staggered case - all groups} have been considered in the literature. For example, \citet{callaway_difference--differences_2020} consider a relaxation of Assumption \ref{asm: pt in staggered case - all groups} that imposes (\ref{eqn: pt for groups}) only for years after some units are treated: 

\let\origtheasm\theasm
\edef\oldasm{\the\numexpr\value{asm}}
\setcounter{asm}{0}
\renewcommand{\theasm}{\oldasm.\alph{asm}}

\begin{asm}[Parallel trends for staggered setting - post-treatment only] \label{asm: pt in staggered case - post-treatment}
\begin{equation*}
\expe{ Y_{i,t}(\infty) - Y_{i,t'}(\infty) \,|\, G_i = g } =  \expe{ Y_{i,t}(\infty) - Y_{i,t'}(\infty) \,|\, G_i = g' }.
\end{equation*} 
for all $t,t' \geq g_{min}-1$, where $g_{min} = \min \mathcal{G}$ is the first period where a unit is treated. 
\end{asm}
\let\theasm\origtheasm
\setcounter{asm}{\the\numexpr\value{asm}+3}

\noindent This would require, for example, that groups of states that expanded Medicaid at different times have parallel trends in $Y_{i,t}(\infty)$ after Medicaid expansion began, but does not necessarily impose parallel trends in the pre-treatment period. Likewise, several papers, including \citet{callaway_difference--differences_2020} and \citet{sun_estimating_2020}, consider versions that impose (\ref{eqn: pt for groups}) only for groups that are eventually treated, and not for the never-treated group (i.e. excluding $g=\infty$). This would impose, for example, parallel trends among states that eventually expanded Medicaid, but not between eventually-adopting and never-adopting states. There are tradeoffs between the different forms of Assumption \ref{asm: pt in staggered case - all groups}: imposing parallel trends for all groups and all periods is a stronger assumption and thus may be less plausible; on the other hand, it may allow one to obtain more precise estimates.\footnote{If all units are eventually treated, then imposing parallel trends only among treated units also limits the number of periods for which the ATT is identified.}  We return to these tradeoffs in our discussion of different estimators for the staggered case below.\footnote{In this paper, we specify the parallel trends assumption based on groups defined by the treatment starting date. It is also possible to adopt alternative definitions of parallel trends using groups that are more disaggregated than our $G$. For instance, one could impose parallel trends for all pairs of states, rather than for groups of states with the same treatment start date. Using a more disaggregated definition of a group strengthens the parallel trends assumption, but could potentially enable more efficient estimators. We focus on the group-level version of parallel trends for simplicity.}

\paragraph{No anticipation.} The no-anticipation assumption from the canonical model also extends in a straightforward way to the staggered setting. Intuitively, it imposes that if a unit is untreated in period $t$, their outcome does not depend on what time period they will be treated in the future --- that is, units do not act on the knowledge of their future treatment date before treatment starts.
\begin{asm}[Staggered no anticipation assumption] \label{asm: no anticipation - staggered}
$Y_{i,t}(g) = Y_{i,t}(\infty)$ for all $i$ and $t<g$.
\end{asm}

\subsection{Interpreting the estimand of two-way fixed effects models}

Recall that in the simple two-period model, the estimand (population coefficient) of the two-way fixed effects specification (\ref{eqn: TWFE-2-periods}) corresponds with the ATT under the parallel trends and no anticipation assumptions. A substantial focus of the recent literature has been whether the estimand of commonly-used generalizations of this TWFE model to the multi-period, staggered timing case have a similar, intuitive causal interpretation. In short, the literature has shown that the estimand of TWFE specifications in the staggered setting often does not correspond with an intuitive causal parameter even under the natural extensions of the parallel trends and no-anticipation assumptions described above.

\paragraph{Static TWFE.}

We begin with a discussion of the ``static'' TWFE specification, which regresses the outcome on individual and period fixed effects and an indicator for whether the unit $i$ is treated in period $t$, 
\begin{equation}
Y_{i,t} = \alpha_i + \phi_t + D_{i,t} \beta_{post} + \epsilon_{i,t}. \label{eqn: static-TWFE-multiple-periods}
\end{equation}

The static specification yields a sensible estimand when there is no heterogeneity in treatment effects across either time or units. Formally, let $\tau_{i,t}(g) = Y_{i,t}(g) - Y_{i,t}(\infty)$. Suppose that for all units $i$, $\tau_{i,t}(g) = \tau$ whenever $t\geq g$. This imposes that (1) all units have the same treatment effect, and (2) the treatment has the same effect regardless of how long it has been since treatment started. In our ongoing example, this would impose that the effect of Medicaid expansion on insurance coverage is the same both across states and across time. Then, under a suitable generalization of the parallel trends assumption (e.g. Assumption \ref{asm: pt in staggered case - all groups}) and no anticipation assumption (Assumption \ref{asm: no anticipation - staggered}), the population regression coefficient $\betapost$ from (\ref{eqn: static-TWFE-multiple-periods}) is equal to $\tau$.

Issues arise with the static specification, however, when there is heterogeneity of treatment effects over time, as shown in \citet{borusyak_revisiting_2016}, \citet{de_chaisemartin_two-way_2020}, and  \citet{goodman-bacon_difference--differences_2018}, among others. Suppose first that there is heterogeneity in time since treatment only. That is, $\tau_{i,t}(g) = \sum_{s\geq0} \tau_{s} 1[t-g = s]$, so all units have treatment effect $\tau_s$ in the $s$-th period after they receive treatment. In this case, $\betapost$ corresponds with a potentially non-convex weighted average of the parameters $\tau_s$, i.e. $\betapost = \sum_s \omega_s \tau_s$, where the weights $\omega_s$ sum to 1 but may be negative. The possibility of negative weights is concerning because, for instance, all of the $\tau_s$ could be positive and yet the coefficient $\beta_{post}$ may be negative! In particular, longer-run treatment effects will often receive negative weights. Thus, for example, it is possible that the effect of Medicaid expansion on insurance coverage is positive and grows over time since the expansion, and yet $\beta_{post}$ in (\ref{eqn: static-TWFE-multiple-periods}) will be negative. More generally, if treatment effects vary across both time and units, then $\tau_{i,t}(g)$ may get negative weight in the TWFE estimand for some combinations of $t$ and $g$.\footnote{We focus in this section on decompositions of the static TWFE model in a standard, sampling-based framework. \citet{athey_design-based_2018} study the static specification in a finite-sample randomization-based framework.}

\citet{goodman-bacon_difference--differences_2018} provides some helpful intuition to understand this phenomenon. He shows that $\betahatpost$ can be written as a convex weighted average of differences-in-differences comparisons between pairs of units and time periods in which one unit changed its treatment status and the other did not. Counterintuitively, however, this decomposition includes difference-in-differences that use as a ``control'' group units who were treated in earlier periods. For example, in 2016, a state that first expanded Medicaid in 2014 might be used as the ``control group'' for a state that first adopted Medicaid in 2016. Hence, an early-treated unit can get negative weights if it appears as a ``control'' for many later-treated units. This decomposition further highlights that $\betapost$ may not be a sensible estimand when treatment effects differ across either units or time, because of its inclusion of these ``forbidden comparisons''.\footnote{To the best of our knowledge, the phrase ``forbidden comparisons'' was introduced in \citet{borusyak_revisiting_2016}.}   

We now give some more mathematical intuition for why weighting issues arise in the static specification with heterogeneity. From the Frisch-Waugh-Lovell theorem, the coefficient $\betapost$ from (\ref{eqn: static-TWFE-multiple-periods}) is equivalent to the coefficient from a univariate regression of $Y_{i,t}$ on $D_{i,t} - \widehat{D}_{i,t}$, where $\widehat{D}_{i,t}$ is the predicted value from a regression of $D_{i,t}$ on the other right-hand side variables in (\ref{eqn: static-TWFE-multiple-periods}), $D_{i,t} = \tilde\alpha_i + \tilde\phi_t + u_{i,t}$. However, a well-known issue with OLS with binary outcomes is that its predictions may fall outside the unit interval. If the predicted value $\widehat{D}_{i,t}$ is greater than 1, then $D_{i,t} - \widehat{D}_{i,t}$ will be negative even when a unit is treated, and thus that unit's outcome will get negative weight in $\betahatpost$. To see this more formally, we can apply the formula for univariate OLS coefficients to obtain that

\begin{equation}
\betahatpost = \dfrac{\sum_{i,t} (D_{i,t} - \widehat{D}_{i,t}) Y_{i,t}}{\sum_{i,t} (D_{i,t} - \widehat{D}_{i,t})^2}.     
\end{equation}

\noindent The denominator is positive, and so the weight that $\betahatpost$ places on $Y_{i,t}$ is proportional to $D_{i,t} - \widehat{D}_{i,t}$. Thus, if $D_{i,t}=1$ and $D_{i,t} - \widehat{D}_{i,t} <0$, then $\betahatpost$ will be decreasing in $Y_{i,t}$ even though unit $i$ is treated at period $t$. But because $Y_{i,t} = Y_{i,t}(\infty) + \tau_{i,t}(g)$, it follows that $\tau_{i,t}(g)$ gets negative weight in $\betahatpost$. 

These negative weights will tend to arise for early-treated units in periods late in the sample. To see why this is the case, we note that some algebra shows that $\widehat{D}_{i,t} = \overline{D}_i + \overline{D}_t - \overline{D}$, where $\overline{D}_i = {T}^{-1} \sum_{t} D_{i,t}$ is the time average of $D$ for unit $i$, $\overline{D}_{t} = {N}^{-1} \sum_{i} D_{i,t}$ is the cross-sectional average of $D$ for period $t$, and $\overline{D} = {(NT)}^{-1} \sum_{i,t} D_{i,t} $ is the average of $D$ across both periods and units. It follows that if we have a unit that has been treated for almost all periods ($\overline{D}_i \approx 1$) and a period in which almost all units have been treated ($\overline{D}_t \approx 1$), then  $\widehat{D}_{i,t} \approx 2 - \overline{D}$, which will be strictly greater than 1 if there is a non-substantial fraction of non-treated units in some period $(\overline{D} < 1)$. We thus see that $\betahatpost$ will tend to put negative weight on $\tau_{i,t}$ for early-adopters in late periods within the sample. This decomposition makes clear that the static OLS coefficient $\betahatpost$ is not aggregating natural comparisons of units, and thus will not produce a sensible estimand when there is arbitrary heterogeneity. When treatment effects are homogeneous -- i.e. $\tau_{i,t}(g) \equiv \tau$ -- the negative weights on $\tau$ for some units cancel out the positive weights on other units, and thus $\betapost$ recovers the causal effect under a suitable generalization of parallel trends. 

\paragraph{Dynamic TWFE.} Next, we turn our attention to the ``dynamic specification'' that regresses the outcome on individual and period fixed effects, as well as dummies for time relative to treatment

\begin{equation}
Y_{i,t} = \alpha_i + \phi_t + \sum_{r\neq0} 1[R_{i,t} = r] \beta_{r} + \epsilon_{i,t}, \label{eqn:dynamic-twfe-multiple-periods}
\end{equation}
\noindent where $R_{i,t} = t - G_i +1$ is the time relative to treatment (e.g. $R_{i,t}=1$ in the first treated period for unit $i$), and the summation runs over all possible values of $R_{it}$ except for 0.

Unlike the static specification, the dynamic specification yields a sensible causal estimand when there is heterogeneity only in time since treatment. In particular, the results in \citet{borusyak_revisiting_2016} and \citet{sun_estimating_2020} imply that if $\tau_{i,t}(g) = \sum_{s\geq0} \tau_{s} 1[t-g = s]$, so all units have treatment effect $\tau_s$ in the $s$-th period after treatment, then $\beta_s = \tau_s$ under suitable generalizations of the parallel trends and no anticipation assumptions, such as Assumptions \ref{asm: pt in staggered case - all groups} and \ref{asm: no anticipation - staggered}.\footnote{We note that the homogeneity assumption can be relaxed so that all adoption cohorts have the same expected treatment effect, i.e. $\expe{ \tau_{i,g+s}(g) \,|\, G = g} \equiv \tau_s$ for all $s$ and $g$. Additionally, these results assume that all possible relative time indicators are included. As discussed in \citet{sun_estimating_2020}, \citet{baker_how_2021}, and \citet{schmidheiny_event_2020}, among others, problems may arise if one ``bins'' endpoints (e.g. includes a dummy for 5$+$ years since treatment).} Thus, specification (\ref{eqn:dynamic-twfe-multiple-periods}) will yield sensible estimates for the dynamic effect of Medicaid expansion if the effect $r$ years after Medicaid expansion is the same (on average) regardless of what year the state initially expanded coverage (for each $r=1,2,...$).

\citet{sun_estimating_2020} show, however, that when there are heterogeneous dynamic treatment effects across adoption cohorts, the coefficients from specification (\ref{eqn:dynamic-twfe-multiple-periods}) become difficult to interpret. Thus, for example, problems may arise if the average treatment effect in the first year after adoption is different for states that adopted Medicaid in 2014 as it is for states that adopted in 2015.\footnote{That is, if the effect in 2015 for the 2014 adoption cohort is different from the effect in 2016 for the 2015 adoption cohort.} There are two issues. First, as with the ``static'' regression specification above, the coefficient $\beta_r$ may put \textit{negative} weight on the treatment effect $r$ periods after treatment for some units. Thus, for example, the treatment effect for some states two years after Medicaid expansion may enter $\beta_2$ negatively. Second, the coefficient $\beta_r$ can put non-zero weight on treatment effects at lags $r'\neq r$, so there is cross-lag ``contamination.'' Thus, for example, the coefficient $\beta_2$ may be influenced by the treatment effect for some states three periods after Medicaid expansion. 

Like the static specification, the dynamic specification thus fails to yield sensible estimates of dynamic causal effects under heterogeneity across cohorts. The derivation of this result is mathematically more complex, and so we do not pursue it here. The intuition is that, as in the static case, the dynamic OLS specification does not aggregate natural comparisons of units and includes ``forbidden comparisons'' between sets of units both of which have already been treated. An important implication of the results derived by \citet{sun_estimating_2020} is that if treatment effects are heterogeneous, the ``treatment lead'' coefficients from (\ref{eqn:dynamic-twfe-multiple-periods}) are not guaranteed to be zero even if parallel trends is satisfied in all periods (and vice versa), and thus evaluation of pre-trends based on these coefficients can be very misleading. 

\subsubsection{Diagnostic approaches\label{subsec: diagnostic approaches}}

Several recent papers introduce diagnostic approaches for understanding the extent of the aggregation issues under staggered treatment timing, with a focus on the static specification (\ref{eqn: static-TWFE-multiple-periods}). \citet{de_chaisemartin_two-way_2020} propose reporting the number/fraction of group-time ATTs that receive negative weights, as well as the degree of heterogeneity in treatment effects that would be necessary for the estimated treatment effect to have the ``wrong sign.'' \citet{goodman-bacon_difference--differences_2018} proposes reporting the weights that $\betahatpost$ places on the different 2-group, 2-period difference-in-differences, which allows one to evaluate how much weight is being placed on ``forbidden'' comparisons of already-treated units and how removing the comparisons would change the estimate. \citet{jakiela_simple_2021} proposes evaluating both whether TWFE places negative weights on some treated units and whether the data rejects the constant treatment effects assumption.

\subsection{New Estimators For Staggered Timing \label{subsec: staggered estimators}}

Several recent papers have proposed alternative estimators that more sensibly aggregate heterogeneous treatment effects in settings with staggered treatment timing. The derivation of each of these estimators follows a similar logic to the derivation of the DiD estimator in the motivating example in Section \ref{sec: basic model}. We begin by specifying a causal parameter of interest (analogous to the ATT $\tau_2$). With the help of the (generalized) parallel trends and no-anticipation assumptions, we can infer the counterfactual outcomes for treated units using trends in outcomes for an appropriately chosen ``clean'' control group of untreated units. This allows us to express the target parameter in terms of identified expectations, analogous to equation (\ref{eqn: identification-tau2-2-periods}). Finally, we replace population expectations with sample averages to form an estimator of the target parameter.

\paragraph{The Callaway and Sant'Anna estimator.} We first describe in detail the approach taken by \citet{callaway_difference--differences_2020}, and then discuss the connections to other approaches. They consider as a building block the group-time average treatment effect on the treated, $ATT(g,t) = \expe{Y_{i,t}(g) - Y_{i,t}(\infty) \,|\, G_{i} = g}$, which gives the average treatment effect at time $t$ for the cohort first treated in time $g$. For example $ATT(2014,2016)$ would be the average treatment effect in 2016 for states who first expanded Medicaid in 2014. They then consider identification and estimation under generalizations of the parallel trends assumption to the staggered setting.\footnote{\citet{callaway_difference--differences_2020} also consider generalizations where the parallel trends assumption holds only conditional on covariates. We discuss this extension in Section \ref{subsec: conditional pt} below, but focus for now on the case without covariates.} Intuitively, under the staggered versions of the parallel trends and no anticipation assumptions, we can identify $ATT(g,t)$ by comparing the expected change in outcome for cohort $g$ between periods $g-1$ and $t$ to that for a control group not-yet treated at period $t$. Formally, under Assumption \ref{asm: pt in staggered case - post-treatment},
\begin{equation*}
ATT(g,t) = \expe{Y_{i,t} - Y_{i,g-1} \,|\, G_i = g} - \expe{Y_{i,t} - Y_{i,g-1} \,|\, G_i = g'}, \text{ for any } g'>t    
\end{equation*}

\noindent which can be viewed as the multi-period analog of the identification result in equation (\ref{eqn: identification-tau2-2-periods}). Since this holds for any comparison group $g'>t$, it also holds if we average over some set of comparisons $\mathcal{G}_{comp}$ such that $g'>t$ for all $g' \in \mathcal{G}_{comp}$,
\begin{equation*}
ATT(g,t) = \expe{Y_{i,t} - Y_{i,g-1} \,|\, G_i = g} - \expe{Y_{i,t} - Y_{i,g-1} \,|\, G_i \in \mathcal{G}_{comp}}.  
\end{equation*}

We can then estimate $ATT(g,t)$ by replacing expectations with their sample analogs,

\begin{equation}
\widehat{ATT}(g,t) = \frac{1}{N_g} \sum_{i \,:\, G_i =g} [Y_{i,t} - Y_{i,g-1}] - \frac{1}{N_{\mathcal{G}_{comp}}} \sum_{i \,:\, G_i \in \mathcal{G}_{comp}} [Y_{i,t} - Y_{i,g-1}]. \label{eqn: callaway sample analog}   
\end{equation}

\noindent Specifically, \citet{callaway_difference--differences_2020} consider two options for $\mathcal{G}_{comp}$. The first uses only never-treated units ($\mathcal{G}_{comp} = \{\infty\}$) and the second uses all not-yet-treated units ($\mathcal{G}_{comp} = \{g' : g' > t\}$).\footnote{We note that if the never-treated units are not included in the comparison group (i.e. $\infty \not\in \mathcal{G}_{comp}$), then one can rely on a weaker version of Assumption \ref{asm: pt in staggered case - post-treatment} that excludes the never-treated group.} When there are a relatively small number of periods and treatment cohorts, reporting $\widehat{ATT}(g,t)$ for all relevant $(g,t)$ may be reasonable.  

When there are many treated periods and/or cohorts, however, reporting all the $\widehat{ATT}(g,t)$ may be cumbersome, and each one may be imprecisely estimated. Thankfully, the method described above extends easily to estimating any weighted average of the $ATT(g,t)$. For instance, we may be interested in an ``event-study'' parameter that gives the (weighted) average of the treatment effect $l$ periods after adoption across different adoption cohorts, 

\begin{equation}
ATT_{l}^{w} = \sum_g w_g ATT(g, g+l). \label{eqn: average att at lag l}    
\end{equation}

\noindent The weights $w_g$ could be chosen to weight different cohorts equally, or in terms of their relative frequencies in the treated population. It is straightforward to form estimates for $ATT^w_{l}$ by averaging the estimates $\widehat{ATT}(g,t)$ discussed above. We refer the reader to \citet{callaway_difference--differences_2020} for a discussion of a variety of other weighted averages that may be economically relevant. Inference is straightforward using either the delta method or a bootstrap, as described in \citet{callaway_difference--differences_2020}.

This alternative approach to estimation has two primary advantages over standard static or dynamic TWFE regressions. The first is that it provides sensible estimands even under arbitrary heterogeneity of treatment effects. By sensible we mean both that the approach avoids negative weighting, but also that the weighting of effects across cohorts is specified by the researcher (e.g. proportional to cohort size) rather than determined by OLS (i.e. proportional to the variance of the treatment indicator). The second advantage is that it makes transparent exactly which units are being used as a control group to infer the unobserved potential outcomes. This contrasts with standard TWFE models, which we have seen make unintuitive comparisons under staggered timing.  

\paragraph{Imputation estimators.} \citet{borusyak_revisiting_2021} introduce a related approach which they refer to as an imputation estimator (see, also, \citet{gardner_two-stage_2021}, \citet{liu_practical_2021} and \citet{Wooldridge2021a} for similar proposals). Specifically, they fit a TWFE regression, $Y_{i,t}(\infty) = \alpha_i + \lambda_t + \epsilon_{i,t}$, using observations only for units and time periods that are not-yet-treated. They then infer the never-treated potential outcome for each treated unit using the predicted value from this regression, $\widehat{Y}_{i,t}(\infty)$. This provides an estimate of the treatment effect for each treated unit, $Y_{i,t} - \widehat{Y}_{i,t}(\infty)$, and these individual-level estimates can be aggregated to form estimates of summary parameters like the $ATT(g,t)$ described above. These approaches yield valid estimates when parallel trends holds for all groups and time periods and there is no anticipation (Assumptions \ref{asm: pt in staggered case - all groups} and \ref{asm: no anticipation - staggered}). 

\paragraph{Comparison of CS and BJS approaches.} How does the approach proposed by \citet[][CS]{callaway_difference--differences_2020} compare to that proposed by \citet[][BJS]{borusyak_revisiting_2021}? For simplicity, it is instructive to consider a simple non-staggered setting where there are three periods ($t=1,2,3$) and units are either treated in period 3 or never-treated ($\mathcal{G} = \{3,\infty\}$). In this case, the CS estimator for the treated group in period 3 (i.e. $ATT(3,3)$) is simply a DiD comparing the treated/untreated units between periods 2 and 3,
$$\widehat{ATT}(3,3) = \underbrace{(\bar{Y}_{3,3} - \bar{Y}_{3,\infty})}_{\text{Diff at $t=3$}} - \underbrace{(\bar{Y}_{2,3} - \bar{Y}_{2,\infty})}_{\text{Diff at $t=2$}} ,$$

\noindent where $\bar{Y}_{t,g}$ is the average outcome in period $t$ for units with $G_i =g$. By contrast, the BJS estimator runs a similar DiD, except instead of using period 2 as the baseline, the BJS estimator uses the \emph{average} outcome prior to treatment (across periods 1 and 2),

$$\widehat{ATT}_{BJS}(3,3) = \underbrace{(\bar{Y}_{3,3} - \bar{Y}_{3,\infty})}_{\text{Diff at $t=3$}} - \underbrace{(\bar{Y}_{pre,3} - \bar{Y}_{pre,\infty})}_{\text{Avg Diff in Pre-Periods}} ,$$
\noindent where $\bar{Y}_{pre,g} = \frac{1}{2}(\bar{Y}_{1,g} + \bar{Y}_{2,g})$ is the average outcome for cohort $g$ across the two pre-treatment periods. Thus, we see that the key difference between the CS and BJS estimators is that CS makes all comparisons relative to the \emph{last pre-treament period}, whereas BJS makes comparisons relative to the \emph{average of the pre-treatment periods}. This primary difference in how the two approaches use pre-treatment periods extends beyond this simple case to settings with staggered timing, although the math becomes substantially more complicated in the staggered case (and thus we do not pursue it).

What are the pros and cons of using the last pre-treatment period versus the average of the pre-treatment periods? In general, there will be tradeoffs between efficiency and the strength of the identifying assumption. On the one hand, averaging over multiple pre-treatment periods can increase precision. Indeed, BJS prove that when Assumption \ref{asm: pt in staggered case - all groups} holds, their estimator is efficient under homoskedasticity and serially uncorrelated errors; see also \citet{Wooldridge2021a}. Although these ideal conditions are unlikely to be satisfied exactly, it does suggest that their estimator will tend to be more efficient than CS when the outcome is not too heteroskedastic or serially correlated.\footnote{By contrast, note that if $Y_{i,t}(0)$ follows a random walk, then $Y_{i,3}(0)$ is independent of $Y_{i,1}(0)$ conditional on $Y_{i,2}(0)$, and thus it is efficient to ignore the earlier pre-treatment periods as CS does.} On the other hand, the two approaches require different identifying assumptions: in the simple example above, CS only relies on parallel trends between periods 2 and 3, whereas BJS relies on parallel trends for all three periods.\footnote{Or more precisely, between the average outcome in periods 1 and 2, and period 3. See also \citet{Marcus2021} for a discussion about different parallel trends assumptions.} More generally, the BJS approach imposes parallel trends for all groups and time periods (Assumption \ref{asm: pt in staggered case - all groups}), whereas the CS approach only relies on post-treatment parallel trends (Assumption \ref{asm: pt in staggered case - post-treatment}). Relying on parallel trends over a longer time horizon may lead to larger biases if the parallel trends assumption holds only approximately: for example, if the average untreated potential outcome is increasing faster among treated units than untreated units over time, then the violation of parallel trends is larger when we compare periods farther apart, and thus the BJS estimator using periods 1 and 2 as the comparison will have larger bias than the CS estimator using only period 2; see \citet{roth_should_2018} and \citet{de_chaisemartin_two-way-survey_2021} for additional discussion. Thus, the BJS estimator may be preferable in settings where the outcome is not too serially correlated and the researcher is confident in parallel trends across all periods; whereas the CS estimator may be preferred in settings where serial correlation is high or the researcher is concerned about the validity of parallel trends over longer time horizons.\footnote{We also note that the BJS and CS estimators incorporate covariates differently. BJS adjust linearly for covariates, where CS consider more general adjustments as described in Section \ref{subsec: conditional pt}.}

\paragraph{Other related approaches.} Several other recent papers propose similar estimation strategies to those described above --- although with some subtle differences in how they construct the control group and the weights they place on different cohorts/time periods. \citet{de_chaisemartin_two-way_2020} propose an estimator that can be applied when treatment turns on and off (see Section \ref{subsec: related timing extensions} below), but in the context of the staggered setting here corresponds with the Callaway and Sant'Anna estimator for $ATT^w_0$ and a particular choice of weights. \citet{sun_estimating_2020} propose an estimator that takes the form (\ref{eqn: callaway sample analog}) but uses either the never-treated units (if they exist) or the last-to-be-treated units as the comparison ($\mathcal{G}_{comp} = \{\max_i G_i\}$), rather than the not-yet-treated. \citet{Marcus2021} propose a recursive estimator that more efficiently exploits the identifying assumptions in \citet{callaway_difference--differences_2020}. See, also, \citet{ImaiKim(19)} and \citet{strezhnev2018semiparametric} for closely related ideas. Another related approach is to run a stacked regression where each treated unit is matched to `clean' (i.e. not-yet-treated) controls and there are separate fixed effects for each set of treated units and its control, as in \citet{cengiz_effect_2019} among others. \citet{gardner_two-stage_2021} shows that this approach estimates a convex weighted average of the $ATT(g,t)$ under parallel trends and no anticipation, although the weights are determined by the number of treated units and variance of treatment within each stacked event, rather than by economic considerations.

\subsection{Further extensions to treatment timing/assignment \label{subsec: related timing extensions}}
Our discussion so far has focused on the case where there is a binary treatment that is adopted at a particular date and remains on afterwards. Several recent papers have studied settings with more complicated forms of treatment assignment. We briefly highlight a few of the recent contributions, and refer the reader to the review in \citet{de_chaisemartin_two-way-survey_2021} for more details. 

\citet{de_chaisemartin_two-way_2020} and \citet{ImaiKim(19)} consider settings where units are treated at different times, but do not necessarily require that treatment is an absorbing state. Their estimators intuitively compare changes in outcomes for units whose treatment status changed to other units whose treatment status remained constant over the same periods. This approach yields an interpretable causal effect under generalizations of the parallel trends assumption and and an additional ``no carryover'' assumption that imposes that the potential outcomes depend only on current treatment status and not on the full treatment history. We note that, as described in \citet{bojinov_panel_2020}, the no carryover assumption may be restrictive in many settings --- for example, if the treatment is a raise in the minimum wage and the outcome is employment, then the no carryover assumption requires that employment in period $t$ depends only on whether the minimum wage was raised in period $t$ and not on the history of minimum wage changes. Recent work has begun to relax the no carryover assumption: one example is \citet{de_chaisemartin_difference--differences_2022}, who allow potential outcomes to depend on the full path of treatments, and instead impose a stronger parallel trends assumption that requires parallel trends in untreated potential outcomes regardless of a unit's path of treatment. 

Other work has considered DiD settings with non-binary treatments. \citet{de_chaisemartin_fuzzy_2018} study ``fuzzy'' DiD settings in which all groups are treated in both time periods, but the proportion of units exposed to treatment increases in one group but not in the other. Finally, \citet{de_chaisemartin_two-way_2021} and \citet{callaway_difference--differences_2021} study settings with multi-valued or continuous treatments.

\subsection{Recommendations}

The results discussed above show that while conventional TWFE specifications make sensible comparisons of treated and untreated units in the canonical two-period DiD setting, in the staggered case they typically make ``forbidden comparisons'' between already-treated units. As a result, treatment effects for some units and time periods receive negative weights in the TWFE estimand. In extreme cases, this can lead the TWFE estimand to have the ``wrong sign'' --- e.g., the estimand may be negative even if all the treatment effects are positive. Even if the weights are not so extreme as to create sign reversals, it may nevertheless be difficult to interpret which comparisons the TWFE estimator is making, as the ``control group'' is not transparent, and the weights it chooses are unlikely to be those most relevant for economic policy.

In our view, if the researcher is not willing to impose assumptions on treatment effect heterogeneity, the most direct remedy for this problem is to use the methods discussed in Section \ref{subsec: staggered estimators} that explicitly specify the comparisons to be made between treatment and control groups, as well as the desired weights in the target parameter. These methods allow one to estimate a well-defined causal parameter (under parallel trends), with transparent weights and transparent comparison groups (e.g. not-yet-treated or never-treated units). This approach, in our view, provides a more complete solution to the problem than the diagnostic approaches discussed in Section \ref{subsec: diagnostic approaches}. Although it is certainly valuable to have a sense of the extent to which conventional TWFE specifications are making bad comparisons, eliminating these undesirable comparisons seems to us a better approach than diagnosing the extent of the issue. Using a TWFE specification may be justified for efficiency reasons if one is confident that treatment effects are homogeneous, but researchers will often be unwilling to restrict treatment effect heterogeneity. 

The question of which of the many heterogeneity-robust DiD methods discussed in Section \ref{subsec: staggered estimators} to use is trickier. As described above, the estimators differ in who they use as the comparison group (e.g. not-yet-treated versus never-treated) as well as the pre-treatment time periods used in the comparisons (e.g. the whole pre-treatment period versus the final untreated period). This leads to some tradeoffs between efficiency and the strength of the parallel trends assumption needed for identification, as highlighted in the comparison of BJS and CS above. The best estimator to use will therefore depend on the context --- particularly, on which group is the most sensible comparison, and how confident the researcher is in parallel trends for all periods. Nevertheless, it is our practical experience that the various heterogeneity-robust DiD estimators typically (although not always) produce similar answers. The first-order consideration is therefore to use an approach that makes clear what the target parameter is and which groups are being compared for identification. Thankfully, there are now statistical packages that make implementing (and comparing) the results from these estimators straightforward in practice (see Table \ref{tbl:packages}).

We acknowledge that these new methods may initially appear complicated to researchers accustomed to analyzing seemingly simple regression specifications such as (\ref{eqn: static-TWFE-multiple-periods}) or (\ref{eqn:dynamic-twfe-multiple-periods}). However, while traditional TWFE regressions are easy to \textit{specify}, as discussed above they are actually quite difficult to \textit{interpret}, since they make complicated and unintuitive comparisons across groups. By contrast, the methods that we recommend have a simple interpretation using a coherent comparison group. And while more complex to express in regression format, they can be viewed as simple aggregations of comparisons of group means. We suspect that once researchers gain experience using the newer heterogeneity-robust DiD methods, they will not seem so scary after all!

\section{Relaxing or allowing the parallel trends assumption to be violated\label{sec: pt violated}}

A second strand of the literature has focused on the possibility that the canonical parallel trends assumption may not hold exactly. Approaches to this problem include relaxing the parallel trends assumption to hold only conditional on covariates, testing for pre-treatment violations of the parallel trends assumption, and various tools for robust inference and sensitivity analysis that explore the possibility that parallel trends may be violated in certain ways. 

\subsection{Why might parallel trends be violated?}

The canonical parallel trends assumption requires that the mean outcome for the treated group would have evolved in parallel with the mean outcome for the untreated group if the treatment had not occurred. As discussed in Section \ref{sec: basic model}, this allows for confounding factors that affect treatment status, but these must have a constant additive effect on the mean outcome. 

In practice, however, we will often be unsure of the validity of the parallel trends assumption for several reasons. First, there will often be concern about \textit{time-varying} confounding factors. For example, Democratic-leaning states were more likely to adopt Medicaid expansions but also might be subject to different time-varying macro-economic shocks. A second concern relates to the potential sensitivity of the parallel trends assumption to the chosen function form of the outcome. If parallel trends holds using the outcome measured in levels, $Y_{i,t}(0)$, then it will generally not be the case that it holds for the outcomes measured in logs $log(Y_{i,t}(0))$ (or vice versa). Indeed, \citet{roth_when_2021} show that parallel trends can hold for all monotonic transformations of the outcome $g(Y_{i,t}(0))$ essentially only if the population can be divided into two groups, where the first group is as good as randomly assigned between treatment and control, and the second group has the same potential outcome distribution in both periods. Although there are some cases where these conditions may be (approximately) met --- the most prominent of which is random assignment of treatment --- they are likely not to hold in most settings where DiD is used, and thus parallel trends will be sensitive to functional form. It will often not be obvious that parallel trends should hold for the particular functional form chosen for our analysis --- e.g. should we use insurance rates, or log insurance rates? --- and thus we may be skeptical of its validity.

\subsection{Parallel trends conditional on covariates\label{subsec: conditional pt}}

One way to increase the credibility of the parallel trends assumption is to require that it holds only conditional on covariates. Indeed, if we condition on a rich enough set of covariates $X_i$, we may be willing to believe that treatment is nearly randomly assigned conditional on $X_i$. Imposing only parallel trends conditional on $X_i$ gives us an extra degree of robustness, since conditional random assignment can fail so long as the remaining unobservables have a time-invariant additive effect on the outcome. In the Medicaid expansion example, for instance, we may want to condition on a state's partisan lean.

In the canonical model discussed in Section \ref{sec: basic model}, the parallel trends assumption can be naturally extended to incorporate covariates as follows. 
\begin{asm}[Conditional Parallel Trends] \label{asm: conditional-parallel-trends-2-periods}
\begin{equation}
\expe{ Y_{i,2}(0) - Y_{i,1}(0) \,|\, D_i = 1, X_i } = \expe{ Y_{i,2}(0) - Y_{i,1}(0) \,|\, D_i = 0, X_i } \text{ (almost surely) },  \label{eqn: pt conditional on covariates}
\end{equation}
\noindent for $X_i$ a pre-treatment vector of observable covariates. 
\end{asm} 

\noindent For simplicity, we will focus first on the conditional parallel trends assumption in the canonical two-period model, although several papers have also extended this idea to the case of staggered treatment timing, as we will discuss towards the end of this subsection. We furthermore focus our discussion on covariates that are measured prior to treatment and that are time-invariant (although they may have a time-varying impact on the outcome); relevant extensions to this are also discussed below. 

In addition to the conditional parallel trends assumption, we will also impose an overlap condition (a.k.a. positivity condition), which guarantees that for each treated unit with covariates $X_i$, there are at least some untreated units in the population with the same value of $X_i$. This overlap assumption is particularly important for using standard inference procedures \citep{Khan2010}.

\begin{asm}[Strong overlap]\label{asm: strong overlap}
The conditional probability of belonging to the treatment group, given observed characteristics, is uniformly bounded away from one, and the proportion of treated units is bounded away from zero. That is, for some $\epsilon>0$, $P(D_i=1|X_i) < 1-\epsilon$, almost surely, and $\expe{D}> 0.$
\end{asm}

Given the conditional parallel trends assumption, no anticipation assumption, and overlap condition, the ATT conditional on $X_i =x $,
$$\tau_2(x) = \expe{Y_{i,2}(1) - Y_{i,2}(0) | D_i=1, X_i=x},$$
\noindent is identified for all $x$ with $P(D_i=1|X_i=x) > 0$. In particular,

\begin{equation}
\tau_2(x) = \underbrace{\expe{Y_{i,2} - Y_{i,1} \,|\,D_i = 1, X_i = x}}_{\text{ Change for $D_i=1,X_i=x$}} - \underbrace{\expe{Y_{i,2} - Y_{i,1} \,|\, D_i = 0, X_i =x}}_{\text{Change for $D_i = 0, X_i=x$}}.  \label{eqn: identification-tau2-conditional-2-periods}  
\end{equation}

\noindent Note that equation (\ref{eqn: identification-tau2-conditional-2-periods}) is analogous to (\ref{eqn: identification-tau2-2-periods}) in the canonical model, except it conditions on $X_i =x$. Intuitively, among the sub-population with $X_i =x$, we have parallel trends, and so we can take the same steps as in Section \ref{sec: basic model} to infer the conditional ATT for that sub-population. The unconditional ATT can then be identified by averaging $\tau_2(x)$ over the distribution of $X_i$ in the treated population. Using the law of iterated expectations, we have 
$$\tau_2 = \expe{Y_{i,2}(1) - Y_{i,2}(0) |D_i =1 } =  \expe{ \underbrace{\expe{Y_{i,2}(1) - Y_{i,2}(0) |D_i =1, X_i}}_{\tau_2(X_i)} | D_i = 1 }.$$

When $X_i$ is discrete and takes a small number of values --- for example, if $X_i$ is an indicator for whether someone has a college degree -- then estimation is straightforward. We can just run an unconditional DiD for each value of $X_i$, and then aggregate the estimates to form an estimate for the overall ATT, using the delta method or bootstrap for the standard errors. When $X_i$ is either continuously distributed or discrete with a very large number of support points, estimation becomes more complicated, because we will typically not have a large enough sample to do an unconditional DiD within each possible value of $X_i$. Thankfully, there are several available econometric approaches to semi-/non-parametrically estimate the ATT even with continuous covariates. We first discuss the limitations of using TWFE regressions in this setting, and then discuss several alternative approaches.

\paragraph{Standard linear regression.} Given that the TWFE specification (\ref{eqn: TWFE-2-periods}) yielded consistent estimates of the ATT under Assumptions \ref{asm: parallel-trends-2-periods}-\ref{asm: random sampling - two periods} in the canonical DiD model, it may be tempting to augment this specification with controls for a time-by-covariate interaction,
\begin{equation}
Y_{i,t} = \alpha_i + \phi_t + (1[t=2] \cdot D_i) \beta  + (X_i \cdot 1[t=2]) \gamma + \varepsilon_{i,t}, \label{eqn:twfe-with-covariates} 
\end{equation}
\noindent for estimation under conditional parallel trends. Unfortunately, this augmented specification need not yield consistent estimates of the ATT under conditional parallel trends without additional homogeneity assumptions. The intuition is that equation (\ref{eqn:twfe-with-covariates}) implicitly models the conditional expectation function (CEF) of $Y_{i,2}-Y_{i,1}$ as depending on $X_i$ with a constant slope of $\gamma$, regardless of $i$'s treatment status. If there are heterogeneous treatment effects that depend on $X_i$ --- e.g., the ATT varies by age of participants --- the derivative of the CEF with respect to $X_i$ may depend on treatment status $D_i$ as well. In these practically relevant setups, estimates of $\beta$ can be biased for the ATT; see \citet{Meyer1995_jbes} and \citet{abadie_semiparametric_2005} for additional discussion. Fortunately, there are several semi-/non-parametric methods available that allow for consistent estimation of the ATT under conditional parallel trends under weaker homogeneity assumptions.

\paragraph{Regression adjustment.} An alternative approach to allow for covariate-specific trends in DiD settings is the regression adjustment procedure proposed by \citet{Heckman1997} and \citet{heckman_characterizing_1998}. Their main idea exploits the fact that under conditional parallel trends, strong overlap, and no anticipation we can write the ATT as
\begin{eqnarray}
\tau_2 &=& \expe{\left. \, \expe{ Y_{i,2} - Y_{i,1} \,|\, D_i = 1, X_i} - \expe{ Y_{i,2} - Y_{i,1} \,|\, D_i = 0, X_i} \, \right| \, D_i = 1}, \nonumber\\
&=& \expe{ Y_{i,2} - Y_{i,1} \,|\, D_i = 1} -  \expe{\left.\expe{ Y_{i,2} - Y_{i,1} \,|\, D_i = 0, X_i}  \, \right| \, D_i = 1}, \nonumber
\end{eqnarray}
where the second equality follows from the  law of iterated expectations. Thus, to estimate the ATT under conditional parallel trends, one simply needs to estimate the conditional expectation of the outcome among untreated units, and then average these ``predictions'' using the empirical distribution of $X_i$ among treated units. That is, we estimate $\tau_2$ with
\begin{equation}\widehat\tau_2 = \frac{1}{N_1} \sum_{i:D_i=1} \left( (Y_{i,2} - Y_{i,1}) - \widehat{\mathbb{E}}[Y_{i,2} - Y_{i,1} | D_i =0, X_i]\right),\end{equation}
\noindent where $\widehat{\mathbb{E}}[Y_{i,2} - Y_{i,1} | D_i=0, X_i]$ is the estimated conditional expectation function fitted on the control units (but evaluated at $X_i$ for a treated unit). We note that if one uses a linear model for $\widehat{\mathbb{E}}[Y_{i,2} - Y_{i,1} | D_i=0, X_i]$, then this would be similar to a modification of (\ref{eqn:twfe-with-covariates}) that interacts $X_i$ with both treatment group and time dummies, although the two are not quite identical because the outcome regression approach re-weights using the distribution of $X_i$ among units with $D_i =1$. The researcher need not restrict themselves to linear models for the CEF, however, and can use more flexible semi-/non-parametric methods instead. One popular approach in empirical practice is to match each treated unit to a ``nearest neighbor'' untreated unit with similar (or identical) covariate values, and then estimate $\widehat{\mathbb{E}}[Y_{i,2} - Y_{i,1} | D_i=0, X_i]$ using $Y_{l(i)2} - Y_{l(i)1}$, where $l(i)$ is the untreated unit matched to $i$, in which case $\widehat\tau_2$ reduces to the simple DiD estimator between treated units and the matched comparison group. 

The outcome regression approach will generally be consistent for the ATT when the outcome model used to estimate $\widehat{\mathbb{E}}[Y_{i,2} - Y_{i,1} | D_i=0, X_i]$ is correctly specified. Inference can be done using the delta-method for parametric models, and there are also several methods available for semi-/non-parametric models (under some additional regularity conditions), including the bootstrap,  as described in \citet{heckman_characterizing_1998}. Inference is more complicated, however, when one models the outcome evolution of untreated units using a nearest-neighbor approach with a fixed number of matches: the resulting estimator is no longer asymptotically linear and thus standard bootstrap procedures are not asymptotically valid \citep[e.g.,][]{Abadie2006, Abadie2008, Abadie2011, Abadie2012}. Ignoring the matching step can also cause problems, and one therefore needs to use inference procedures that accommodate matching as described in the aforementioned papers.\footnote{Although these nearest-neighbor procedures have been formally justified for cross-sectional data, they are easily adjustable to the canonical 2x2 DiD setup with balanced panel data. We are not aware of formal extensions that allows for unbalanced panel data, repeated cross-sectional data, or more general DiD designs. \citet{abadie_robust_2022} show that, in some cases, clustering at the match level is sufficient when matching is done without replacement.}

\paragraph{Inverse probability weighting} An alternative to modeling the conditional expectation function is to instead model the propensity score, i.e. the conditional probability of belonging to the treated group given covariates, $p(X_i) = P(D_i = 1 | X_i)$. Indeed, as shown by \citet{abadie_semiparametric_2005}, under Assumptions \ref{asm: no anticipation}, \ref{asm: conditional-parallel-trends-2-periods} and \ref{asm: strong overlap}, the ATT is identified using the following inverse probability weighting (IPW) formula:
\begin{eqnarray}
\tau_2&=& \dfrac{\expe{\left(D_i - \dfrac{(1-D_i)p(X_i)}{1-p(X_i)}\right)\left(Y_{i,2}-Y_{i,1}\right)}}{\expe{D_i}}.\label{eqn: IPW_abadie}
\end{eqnarray}

As in the regression adjustment approach, researchers can use the ``plug-in principle'' to estimate the ATT by pluging in an estimate of the propensity score to the equation above. The propensity score model can be estimated using parametric models or semi-/non-parametric models (under suitable regularity conditions). The IPW approach will generally be consistent if the model for the propensity scores is correctly specified. Inference can be conducted using standard tools; see, e.g., \citet{abadie_semiparametric_2005}.


\paragraph{Doubly-robust estimators} The outcome regression and IPW approaches described above can also be combined to form ``doubly-robust'' (DR) methods that are valid if either the outcome model or the propensity score model is correctly specified. Specifically, \citet{santanna_doubly_2020} show that under Assumptions \ref{asm: no anticipation}, \ref{asm: conditional-parallel-trends-2-periods} and \ref{asm: strong overlap}, the ATT is identified as:
\begin{eqnarray}
\tau_2 &=& \expe{\left(\dfrac{D_i}{\expe{D_i}} - \dfrac{\dfrac{(1-D_i) p(X_i)}{1-p(X_i)}}{\expe{\dfrac{(1-D_i)p(X_i)}{1-p(X_i)}}}\right)\left(Y_{i,2}-Y_{i,1} - \expe{ Y_{i,2} - Y_{i,1} \,|\, D_i = 0, X_i} \right)}\label{eqn: DR DiD}.
\end{eqnarray}

As before, one can then estimate the ATT by plugging in estimates of both the propensity score and the CEF. The outcome equation and the propensity score can be modeled with either parametric or semi-/non-parametric methods, and DR methods will generally be consistent if either of these models is correctly specified. In addition, \citet{Chang2020} shows that data-adaptive/machine-learning methods can also be used with DR methods. Standard inference tools can be used as well; see, e.g., \citet{santanna_doubly_2020}. Finally, under some regularity conditions, the DR estimator achieves the semi-parametric efficiency bound when both the outcome and propensity score models are correctly specified \citep{santanna_doubly_2020}.

\paragraph{Extensions to staggered treatment timing:} Although the discussion above focused on DiD setups with two groups and two periods, these different procedures have been extended to staggered DiD setups when treatments are binary and non-reversible. More precisely, \citet{callaway_difference--differences_2020} extend the regression adjustment, IPW and DR procedures above to estimate the family of $ATT(g,t)$'s discussed in Section \ref{subsec: staggered estimators}. They then aggregate these estimators to form different treatment effect summary measures. \citet{Wooldridge2021a} proposes an alternative regression adjustment procedure that is suitable for staggered setups. His proposed estimator differs from the \citet{callaway_difference--differences_2020} regression adjustment estimator as he exploits additional information from pre-treatment periods, which, in turn, can lead to improvements in precision. On the other hand, if these additional assumptions are violated, \citet{Wooldridge2021a}'s estimator may be more biased than \citet{callaway_difference--differences_2020}'s. \citet{de_chaisemartin_two-way_2020, de_chaisemartin_difference--differences_2022} and \citet{borusyak_revisiting_2021} consider estimators which include covariates in a linear manner. 

\paragraph{Caveats.} Throughout, we assume that the covariates $X_i$ were measured prior to the introduction of the intervention and are, therefore, unaffected by it. If $X_i$ can in fact be affected by treatment, then conditioning on it induces a ``bad control'' problem that can induce bias; see \citet{zeldow_confounding_2021} for additional discussion. Similar issues arise if one conditions on time-varying covariates $X_{i,t}$ that can be affected by the treatment. If one is willing to assume that a certain time-varying covariate $W_{i,t}$ is not affected by the treatment, then in principle the entire time-path of the covariate $W_i = (W_{i,1},...,W_{i,T})'$ can be included in the conditioning variable $X_i$, and thus exogenous time-varying covariates can be incorporated similarly to any pre-treatment covariate. See \citet{caetano_difference_2022} for additional discussion of time-varying covariates.

Another important question relates to whether researchers should condition on pre-treatment outcomes. Proponents of including pre-treatment outcomes argue that controlling for lagged outcomes can reduce bias from unobserved confounders \citep{ryan_well-balanced_2018}. It is worth noting when lagged outcomes are included in $X_i$, the conditional parallel trends assumption actually reduces to a conditional mean independence assumption for the untreated potential outcome, since the $Y_{i,1}(0)$ terms on both sides of (\ref{eqn: pt conditional on covariates}) cancel out, and thus we are left with
\begin{equation*}
\expe{ Y_{i,2}(0) \,|\, D_i = 1, X_i } = \expe{ Y_{i,2}(0) \,|\, D_i = 0, X_i } \text{ (almost surely) }.  \label{eqn: conditional unconfoundedness}
\end{equation*}

\noindent Including the lagged outcome in the conditioning variable thus makes sense if one is confident in the conditional unconfoundedness assumption: i.e., if treatment is as good as randomly assigned conditional on the lagged outcome and other elements of $X_i$. This may be sensible in settings where treatment takeup decisions are made on the basis of lagged outcomes. However, it is also important to note that conditioning on lagged outcomes need not necessarily reduce bias, and can in fact exacerbate it in certain contexts. For example, \citet{daw_matching_2018} show that when the treated and comparison groups have different outcome distributions but the same trends, matching the treated and control groups on lagged outcomes selects control units with a particularly large ``shock'' in the pre-treatment period. This can then induce bias owing to a mean-reversion effect, when in fact not conditioning on lagged outcomes would have produced parallel trends. Thus, whether one should include lagged outcomes or not depends on whether the researcher prefers the non-nested assumptions of conditional unconfoundedness (given the lagged outcome) versus parallel trends. See, also \citet{chabe-ferret_analysis_2015}, \citet[][Chapter 5.4]{AngristPischke(09)}, and \citet{Ding2019} for related discussion.

\subsection{Testing for pre-existing trends \label{subsec: pre-trends testing}}

Although conditioning on pre-existing covariates can help increase the plausibility of the parallel trends assumption, researchers typically still worry that there remain unobserved time-varying confounders. An appealing feature of the DiD design is that it allows for a natural plausibility check on the identifying assumptions: did outcomes for the treated and comparison groups (possibly conditional on covariates) move in parallel prior to the time of treatment? It has therefore become common practice to check, both visually and using statistical tests, whether there exist pre-existing differences in trends (``pre-trends'') as a test of the plausibility of the parallel trends assumption.

To fix ideas, consider a simple extension of the canonical non-staggered DiD model in Section \ref{sec: basic model} in which we observe outcomes for an additional period $t=0$ during which no units were treated. (These ideas will extend to the case of staggered treatment or conditional parallel trends). By the no-anticipation assumption, $Y_{i,t} = Y_{i,t}(0)$ for all units in periods $t=0$ and $t=1$. We can thus check whether the analog to the parallel trends assumption held between periods $0$ and $1$ --- that is, is
$$\underbrace{\expe{Y_{i,1} - Y_{i,0} \,|\,D_i =1}}_{\text{Pre-treatment change for $D_i=1$}} - \underbrace{\expe{Y_{i,1} - Y_{i,0} \,|\, D_i = 0}}_{\text{Pre-treatment change for $D_i = 0$}} = 0 ?  $$
\noindent For example, did average insurance rates evolve in parallel for expansion and non-expansion states before either of them expanded Medicaid? In the non-staggered setting, this hypothesis can be conveniently tested using a TWFE specification that includes leads and lags of treatment,
\begin{equation}
Y_{i,t} = \alpha_i + \phi_t + \sum_{r\neq0} 1[R_{i,t}= r] \beta_r + \epsilon_{i,t}, \label{eqn: event-study spec}    
\end{equation}
\noindent where the coefficient on the lead of treatment $\widehat\beta_{-1}$ is given by
$$ \widehat\beta_{-1} = \frac{1}{N_1} \sum_{i:D_i=1} Y_{i,0} - Y_{i,1} - \frac{1}{N_0} \sum_{i:D_i=0} Y_{i,0} - Y_{i,1}.$$
\noindent Testing for pre-treatment trends thus is equivalent to testing the null hypothesis that $\beta_{-1} = 0$. This approach is convenient to implement and extends easily to the case with additional pre-treatment periods and non-staggered treatment adoption. When there are multiple pre-treatment periods, it is common to plot the $\widehat\beta_r$ in what is called as an ``event-study'' plot. If all of the pre-treatment coefficients (i.e., $\widehat{\beta}_r$ for $r<0)$ are insignificant, this is usually interpreted as a sign in favor of the validity of the design, since we cannot reject the null that parallel trends was satisfied in the pre-treatment period.

This pre-testing approach extends easily to settings with staggered adoption and/or conditional parallel trends assumptions. For example, the \citet{callaway_difference--differences_2020} estimator can be used to construct ``placebo'' estimates of $ATT^w_l$ for $l<0$, i.e. the ATT $l$ periods \textit{before} treatment. The estimates $\widehat{ATT}^w_l$ can be plotted for different values of $l$ (corresponding to different lengths of time before/after treatment) to form an event-study plot analogous to that for the non-staggered case. This illustrates that the idea of testing for pre-trends extends easily to the settings with staggered treatment adoption or conditional parallel trends, since the \citet{callaway_difference--differences_2020} can be applied for both of these settings. These results are by no means specific to the \citet{callaway_difference--differences_2020} estimator, though, and event-study plots can be created in a similar fashion using other estimators for either staggered or conditional DiD settings. We caution, however, against using dynamic TWFE specifications like (\ref{eqn: event-study spec}) in settings with staggered adoption, since as noted by \citet{sun_estimating_2020}, the coefficients $\beta_r$ may be contaminated by treatment effects at relative time $r'>0$, so with heterogeneous treatment effects the pre-trends test may reject even if parallel trends holds in the pre-treatment period (or vice versa).

\subsection{Issues with testing for pre-trends \label{subsec: pretesting issues}}
Although tests of pre-existing trends are a natural and intuitive plausibility check of the parallel trends assumption, recent research has highlighted that they also have several limitations. First, even if pre-trends are exactly parallel, this need not guarantee that the post-treatment parallel trends assumption is satisfied. \citet{kahn-lang_promise_2020} give an intuitive example: the average height of boys and girls evolves in parallel until about age 13 and then diverges, but we should not conclude from this that there is a causal effect of bar mitzvahs (which occur for boys at age 13) on children's height!

A second issue is that even if there are pre-existing differences in trends, the tests described above may fail to reject owing to low power \citep{bilinski_seeking_2018, freyaldenhoven_pre-event_2019, kahn-lang_promise_2020, roth_pre-test_2021}. That is, even if there is a pre-existing trend, it may not be significant in the data if our pre-treatment estimates are imprecise.

To develop some intuition for why power may be low, suppose that there is no true treatment effect but there is a pre-existing linear difference in trends between the treatment and comparison groups. Then in the simple example from above, the pre-treatment and post-treatment event-study coefficients will have the same magnitude, $|\beta_{-1}| = |\beta_1|$. If the two estimated coefficients $\widehat\beta_{-1}$ and $\widehat\beta_{1}$ also have the same sampling variance, then by symmetry the probability that the pre-treatment coefficient $\hat\beta_{-1}$ is significant will be the same as the probability that the post-treatment coefficient $\hat\beta_1$ is significant. But this means that a linear violation of parallel trends that would be detected only half the time by a pre-trends test will also lead us to spuriously find a significant treatment effect half the time\footnote{This is the unconditional probability that $\hat\beta_1$ is significant (not conditioning on the result of the pre-test). However, if $\hat\beta_{1}$ and $\hat\beta_{-1}$ are independent, then this is also the probability of finding a significant effect conditional on passing the pre-test.} --- that is, 10 times more often than we expect to find a spurious effect using a nominal 95\% confidence interval! Another intuition for this phenomenon, given by \citet{bilinski_seeking_2018}, is that pre-trends tests reverse the traditional roles of type I and type II error: they set the assumption of parallel trends (or no placebo pre-intervention effect) as the null hypothesis and only ``reject'' the assumption if there is strong evidence against it. This controls the probability of finding a violation when parallel trends holds at 5\% (or another chosen $\alpha$-level), but the probability of failing to identify a violation can be much higher, corresponding to type II error of the test. 

In addition to being concerning from a theoretical point of view, the possibility of low power appears to be relevant in practice: in simulations calibrated to papers published in three leading economics journals, \citet{roth_pre-test_2021} found that linear violations of parallel trends that conventional tests would detect only 50\% of the time often produce biases as large as (or larger than) the estimated treatment effect.

A third issue with pre-trends testing is that conditioning the analysis on ``passing'' a pre-trends test induces a selection bias known as pre-test bias \citep{roth_pre-test_2021}. Intuitively, if there is a pre-existing difference in trends in population, the draws from the DGP in which we fail to detect a significant pre-trend are a selected sample from the true DGP. \citet{roth_pre-test_2021} shows that in many cases, this additional selection bias can exacerbate the bias from a violation of parallel trends.

A final issue with the current practice of pre-trends testing relates to what happens if we do detect a significant pre-trend. In this case, the pre-trends test suggests that parallel trends is likely not to hold exactly, but researchers may still wish to learn something about the treatment effect of interest. Indeed, it seems likely that with enough precision, we will nearly always reject that the parallel trends assumption holds exactly in the pre-treatment period. Nevertheless, we may still wish to learn something about the treatment effect, especially if the violation of parallel trends is ``small'' in magnitude. However, the conventional approach does not make clear how to proceed in this case.  
\subsubsection{Improved diagnostic tools \label{subsec: pretesting better}}
A few papers have proposed alternative tools for detecting pre-treatment violations of parallel trends that take into account some of the limitations discussed above. \citet{roth_pre-test_2021} developed tools to conduct power analyses and calculate the likely distortions from pre-testing under researcher-hypothesized violations of parallel trends. These tools allow the researcher to assess whether the limitations described above are likely to be severe for potential violations of parallel trends deemed economically relevant.

\citet{bilinski_seeking_2018} and \citet{dette_difference--differences_2020} propose ``non-inferiority'' approaches to pre-testing that help address the issue of low power by reversing the roles of the null and alternative hypotheses. That is, rather than test the null that pre-treatment trends are zero, they test the null that the pre-treatment trend is large, and reject only if the data provides strong evidence that the pre-treatment trend is small. For example, \citet{dette_difference--differences_2020} consider null hypotheses of the form $H_0: \max_{r<0} |\beta_r| \geq c$, where $\beta_r$ are the (population) pre-treatment coefficients from regression (\ref{eqn: event-study spec}). This ensures that the test ``detects'' a pre-trend with probability at least $1-\alpha$ when in fact the pre-trend is large (i.e. has magnitude at least $c$).

These non-inferiority approaches are an improvement over standard pre-testing methods, since they guarantee by design that the pre-test is powered against large pre-treatment violations of parallel trends. However, using these approaches does not provide any formal guarantees that ensure the validity of confidence intervals for the treatment effect, the main object of interest. They also do not avoid statistical issues related to pre-testing \citep{roth_pre-test_2021}, and do not provide clear guidance on what to do when the test fails to reject the null of a large pre-trend. This has motivated more formal robust inference and sensitivity analysis approaches that consider inference on the ATT when parallel trends may be violated, which we discuss next.

\subsection{Robust inference and sensitivity analysis \label{subsec: sensitivity analysis}}

\paragraph{Bounds using pre-trends.} \citet{rambachan_honest_2021} propose an approach for robust inference and sensitivity analysis when parallel trends may be violated, building on earlier work by \citet{manski_how_2017}. Their approach attempts to formalize the intuition motivating pre-trends testing: that the counterfactual post-treatment trends cannot be ``too different'' from the pre-trends. To fix ideas, consider the non-staggered treatment adoption setting described in Section \ref{subsec: pretesting issues}. Denote by $\delta_1$ the violation of parallel trends in the first post-treatment period: 
$$\delta_1 = \expe{Y_{i,2}(0) - Y_{i,1}(0) | D_i =1 } -  \expe{Y_{i,2}(0) - Y_{i,1}(0) | D_i =0 } .$$

\noindent This, for example, could be the counterfactual difference in trends in insurance coverage between Medicaid expansion and non-expansion states if the expansions had not occurred. The bias $\delta_1$ is unfortunately not directly identified, since we do not observe the untreated potential outcomes, $Y_{i,2}(0)$, for the treated group. However, by the no anticipation assumption, we can identify the \textit{pre-treatment} analog to $\delta_1$,
$$\delta_{-1} = \expe{Y_{i,0}(0) - Y_{i,1}(0) | D_i =1 } -  \expe{Y_{i,0}(0) - Y_{i,1}(0) | D_i =0 },$$

\noindent which looks at pre-treatment differences in trends between the groups, with $\delta_{-1} = \expe{\betahat_{-1}}$ from the event study regression (\ref{eqn: event-study spec}). For example, $\delta_{-1}$ corresponds to the pre-treatment difference in trends between expansion and non-expansion states.

\citet{rambachan_honest_2021} then consider robust inference under assumptions that restrict the possible values of $\delta_1$ given the value of $\delta_{-1}$ --- or more generally, given $\delta_{-1},...,\delta_{-K}$ if there are $K$ pre-treatment coefficients. For example, one type of restriction they consider states that the magnitude of the post-treatment violation of parallel trends can be no larger than a constant $\bar{M}$ times the maximal pre-treatment violation, i.e. $|\delta_1| \leq \bar{M} \max_{r<0} |\delta_r|$. If $\bar{M}=1$, for example, then this would impose that post-treatment violations of parallel trends are no larger than the largest pre-treatment violation. They also consider restrictions that bound the extent that $\delta_1$ can deviate from a linear extrapolation of the pre-treatment differences in trends. \citet{rambachan_honest_2021} use tools from the partial identification and sensitivity analysis literature \citep{armstrong_optimal_2018, andrews_inference_2019} to construct confidence sets for the ATT that are uniformly valid under the imposed restrictions. These confidence sets take into account the fact that we do not observe the true pre-treatment difference in trends $\delta_{-1}$, only an estimate $\widehat{\beta}_{-1}$. In contrast to conventional pre-trends tests, the \citet{rambachan_honest_2021} confidence sets thus tend to be larger when there is more uncertainty about the pre-treatment difference in trends (i.e. when the standard error on $\widehat{\delta}_{-1}$ is large). 

This approach enables a natural form of sensitivity analysis. For example, a researcher might report that the conclusion of a positive treatment effect is robust up to the value $\bar{M}=2$. This indicates that to invalidate the conclusion of a positive effect, we would need to allow for a post-treatment violation of parallel trends two times larger than the maximal pre-treatment violation. For example, we could potentially say that Medicaid expansion has a significant effect on insurance rates unless we're willing to allow for post-expansion differences in trends that were up to twice as large as the largest difference in trends prior to the expansion. Doing so makes precise exactly what needs to be assumed about possible violations of parallel trends to reach a particular conclusion.

It is worth highlighting that although we've described these tools in the context of non-staggered treatment timing and an unconditional parallel trends assumption, they extend easily to the case of staggered treatment timing and conditional parallel trends as well. Indeed, under mild regularity conditions, these tools can be used anytime the researcher has a treatment effect estimate $\betahatpost$ and a placebo estimate $\betahatpre$, so long as she is willing to bound the possible bias of $\betahatpost$ given the expected value of $\betahatpre$. For example, in the staggered setting, $\betahatpost$ could be an estimate of $ATT^w_l$ for $l>0$ using one of the estimators described in Section \ref{subsec: staggered estimators}, and $\betahatpre$ could be placebo estimates of $ATT^w_{-1}$,...,$ATT^w_{-k}$. See \url{https://github.com/pedrohcgs/CS_RR} for examples on how these sensitivity analyses can be combined with the \citet{callaway_difference--differences_2020} estimator in R. 

\paragraph{Bounds using bracketing.} \citet{ye_negative_2021} consider an alternative partial identification approach where there are two control groups whose trends are assumed to ``bracket'' that of the treatment group. Consider the canonical model from Section \ref{sec: basic model}, and suppose the untreated units can be divided into two control groups, denoted $C_i = a$ and $C_i =b$. For ease of notation, let $C_i = trt$ denote the treated group, i.e. units with $D_i=1$. Let $\Delta(c) = \expe{ Y_{i,2}(0) - Y_{i,1}(0) | C_i =c }$. Instead of the parallel trends assumption, \citet{ye_negative_2021} impose that
\begin{equation}
\min\{ \Delta(a), \Delta(b) \} \leq \Delta(trt) \leq \max\{ \Delta(a), \Delta(b) \},\label{eqn: bracketing}\end{equation}
\noindent so that the trend in $Y(0)$ for the treated group is bounded above and below (``bracketed'') by the minimum and maximum trend in groups $a$ and $b$. An intuitive example where we may have such bracketing is if each of the groups corresponds with a set of industries, and one of the control groups (say group $a$) is more cyclical than the treated group while the other (say group $b$) is less cyclical. If the economy was improving between periods $t=1$ and $t=2$, then we would expect group $a$ to have the largest change in the outcome and group $b$ to have the smallest change; whereas if the economy was getting worse, we would expect the opposite. Under equation (\ref{eqn: bracketing}) and the no anticipation assumption, the ATT is bounded,
$$\expe{Y_{i,2}- Y_{i,1} | D_i=1} - \max\{ \Delta(a), \Delta(b) \} \leq \tau_2 \leq \expe{Y_{i,2} - Y_{i,1} | D_i=1} - \min\{ \Delta(a), \Delta(b) \}.$$
\noindent This reflects that if we knew the true counterfactual trend for the treated group we could learn the ATT exactly, and therefore that bounding this trend means we can obtain bounds on the ATT. \citet{ye_negative_2021} further show how one can construct confidence intervals for the ATT, and extend this logic to settings with multiple periods (but non-staggered treatment timing). See, also, \citet{hasegawa_evaluating_2019} for a related, earlier approach. 

\subsubsection{Other approaches.} \citet{keele_patterns_2019} propose a sensitivity analysis in the canonical two-period DiD model that summarizes the strength of confounding factors that would be needed to induce a particular bias. \citet{freyaldenhoven_visualization_2021} propose a visual sensitivity analysis in which one plots the ``smoothest'' trend though an event-study plot that could rationalize the data under the null of no effect. Finally, \citet{freyaldenhoven_pre-event_2019} propose a GMM-based estimation strategy that allows for parallel trends to be violated when there exists a covariate assumed to be affected by the same confounds as the outcome but not by the treatment itself.

\subsection{Recommendations}

We suspect that in most practical applications of DiD, researchers will not be confident ex ante that the parallel trends assumption holds exactly, owing to concerns about time-varying confounds and sensitivity to functional form. The methods discussed in this setting for relaxing the parallel trends assumption and/or assessing sensitivity to violations of the parallel trends assumption will therefore be highly relevant in most contexts where DiD is applied.

A natural starting point for these robustness checks is to consider whether the results change meaningfully when imposing parallel trends only conditional on covariates. Among the different estimation procedures we discussed, we view doubly-robust procedures as a natural default, since they are valid if either the outcome model or propensity score is well-specified and have desirable efficiency properties. A potential exception to this recommendation arises in settings with limited overlap, i.e., when the estimated propensity score is close to 0 or 1, in which case regression adjustment estimators may be preferred.

Whether one includes covariates into the DiD analysis or not, we encourage researchers to continue to plot ``event-study plots'' that allow for a visual evaluation of pre-existing trends. These plots convey useful information for the reader to assess whether there appears to have been a break in the outcome for the treatment group around the time of treatment. In contexts with a common treatment date, such plots can be created using TWFE specifications like (\ref{eqn: event-study spec}); in contexts with staggered timing, we recommend plotting estimates of $ATT^w_l$ for different values of $l$ using one of the estimators for the staggered setting described in Section \ref{subsec: staggered estimators} to avoid negative weighting issues with TWFE. See Section \ref{subsec: pre-trends testing} for additional discussion. We also refer the reader to \citet{freyaldenhoven_visualization_2021} regarding best-practices for creating such plots, such as displaying simultaneous (rather than pointwise) confidence bands for the path of the event-study coefficients \citep{olea_simultaneous_2019, callaway_difference--differences_2020}.

While event-study plots play an important role in evaluating the plausibility of the parallel trends assumption, we think it is important to appreciate that tests of pre-trends may be underpowered to detect relevant violations of parallel trends, as discussed in Section \ref{subsec: pretesting issues}. \emph{The lack of a significant pre-trend does not necessarily imply the validity of the parallel trends assumption}. At minimum, we recommend that researchers assess the power of pre-trends tests against economically relevant violations of parallel trends, as described in Section \ref{subsec: pretesting better}. 

We also think it should become standard practice for researchers to formally assess the extent to which their conclusions are sensitive to violations of parallel trends. A natural statistic to report in many contexts is the ``breakdown'' value of $\bar{M}$ using the sensitivity analysis in \citet{rambachan_honest_2021} --- i.e. how big would the post-treatment violation of parallel trends have to be relative to the largest pre-treatment violation to invalidate a particular conclusion? We encourage researchers to routinely report the results of the sensitivity analyses described in Section \ref{subsec: sensitivity analysis} alongside their event-study plots.

We also encourage researchers to accompany the formal sensitivity tools with a discussion of possible violations of parallel trends informed by context-specific knowledge. The parallel trends assumption is much more plausible in settings where we expect the trends for the two groups to be similar ex-ante (before seeing the pre-trends). Whenever possible, researchers should therefore provide a justification for why we might expect the two groups to have similar trends. It is also useful to provide context-specific knowledge about the types of confounds that might potentially lead to violations of the parallel trends assumption --- what time-varying factors may have differentially affected the outcome for the treated group? Such discussion can often be very useful for interpreting the results of the formal sensitivity analyses described in Section \ref{subsec: sensitivity analysis}. For example, suppose that a particular conclusion is robust to allowing for violations of parallel trends twice as large the maximum in the pre-treatment period. In contexts where other factors were quite stable around the time of the treatment, this might be interpreted as a very robust finding; on the other hand, if the treatment occurred at the beginning of a recession much larger than anything seen in the pre-treatment period, then a violation of parallel trends of that magnitude may indeed be plausible, so that the results are less robust than we might like. Thus, economic knowledge will be very important in understanding the robustness of a particular result. In our view, the most scientific approach to dealing with possible violations of parallel trends therefore involves a combination of state-of-the-art econometric tools and context-specific knowledge about the types of plausible confounding factors.

\section{Relaxing sampling assumptions\label{sec: inference}}

We now discuss a third strand of the DiD literature, which considers inference under deviations from the canonical assumption that we have sampled a large number of independent clusters from an infinite super-population.

\subsection{Inference procedures with few clusters\label{subsec: model-based-inference}}

As described in Section \ref{sec: basic model}, standard DiD inference procedures rely on researchers having access to data on a large number of treated and untreated clusters. Confidence intervals are then based on the central limit theorem, which states that with independently-sampled clusters, the DiD estimator has an asymptotically normal distribution as the number of treated and untreated clusters grows large. In many practical DiD settings, however, the number of independent clusters (and, in particular, treated clusters) may be small, so that the central limit theorem based on a growing number of clusters may provide a poor approximation. For example, many DiD applications using state-level policy changes may only have a handful of treated states. The central limit theorem may provide a poor approximation with few clusters, even if the number of units within each cluster is large. This is because the standard sampling-based view of clustering allows for arbitrary correlations of the outcome within each cluster, and thus there may be common components at the cluster level (a.k.a. cluster-level ``shocks'') that do not wash out when averaging over many units within the same cluster. Since we only observe a few observations of the cluster-specific shocks, the average of these shocks will generally not be approximately normally distributed. 

\paragraph{Model-based approaches.} Several papers have made progress on the difficult problem of conducting inference with a small number of clusters by modeling the dependence within clusters. These papers typically place some restrictions on the common cluster-level shocks, although the exact restrictions differ across papers. The starting point for these papers is typically a structural equation of the form
\begin{equation}
Y_{ijt} = \alpha_j + \phi_t + D_{jt} \beta + (\nu_{jt} + \epsilon_{ijt}), \label{eqn: static-TWFE-ind-clusters}
\end{equation}
where $Y_{ijt}$ is the (realized) outcome of unit $i$ in cluster $j$ at time $t$, $\alpha_j$ and $\phi_t$ are cluster and time fixed effects, $D_{jt}$ is an indicator for whether cluster $j$ is treated in period $t$, $\nu_{jt}$ is a common cluster-by-time error term, and $\epsilon_{ijt}$ is an idiosyncratic unit-level error term. Here, the ``cluster-level'' error term,  $\nu_{jt}$, induces correlation among units within the same cluster. It is often assumed that $\epsilon_{ijt}$ are $iid$ mean-zero across $i$ and $j$ (and sometimes $t$); see, e.g., \citet{donald_inference_2007}, \citet{Conley2011}, and \citet{Ferman2018}. Letting $Y_{jt} = n_j^{-1} \sum_{i:j(i)=j} Y_{ijt}$ be the average outcome among units in cluster $j$, where $n_j$ is the number of units in cluster $j$, we can take averages to obtain
\begin{equation}
Y_{jt} = \alpha_j + \phi_t + D_{jt} \beta + \eta_{jt}, \label{eqn: static-TWFE-clusters}
\end{equation}
where $\eta_{jt} = \nu_{jt} + n_j^{-1} \sum_{i=1}^{n_j}\epsilon_{ijt}$. Assuming the canonical set-up with two periods where no clusters are treated in period $t=1$ and some clusters are treated in period $t=2$, the canonical DiD estimator at the cluster level is equivalent to the estimated OLS coefficient $\widehat\beta$ from (\ref{eqn: static-TWFE-clusters}), and is given by 
\begin{align*}
\widehat\beta &= \beta + \frac{1}{J_1} \sum_{j:D_j =1} \Delta \eta_{j} - \frac{1}{J_0} \sum_{j:D_j =0} \Delta \eta_{j}\\
& =  \beta + \frac{1}{J_1} \sum_{j:D_j =1} \left(\Delta \nu_{j} + n_j^{-1} \sum_{i=1}^{n_j}\Delta \epsilon_{ij}\right) - \frac{1}{J_0} \sum_{j:D_j =0} \left(\Delta \nu_{j} + n_j^{-1} \sum_{i=1}^{n_j}\Delta \epsilon_{ij}\right) \numberthis \label{eqn: hatbeta-structural-model},
\end{align*}
\noindent where $J_d$ corresponds with the number of \textit{clusters} with treatment $d$, and $\Delta \eta_j = \eta_{j2} - \eta_{j1}$ (and likewise for the other variables).
The equation in the previous display highlights the challenge in this setup: with few clusters, the averages of the cluster level shocks $\Delta \nu_{j}$ among treated and untreated clusters will tend not to be approximately normally distributed, and their variance may be difficult to estimate.

It is worth highlighting that the model described above starts from the structural equation (\ref{eqn: static-TWFE-ind-clusters}) rather than a model where the primitives are potential outcomes as in Section \ref{sec: basic model}. We think that connecting the assumptions on the errors in the structural model (\ref{eqn: static-TWFE-ind-clusters}) to restrictions on the potential outcomes is an interesting open topic for future work. Although a general treatment is beyond the scope of this paper, in Appendix \ref{sec: model based appendix} we show that the errors in the structural model (\ref{eqn: static-TWFE-ind-clusters}) map to primitives based on potential outcomes in the canonical model from Section \ref{sec: basic model}. For the remainder of the sub-section, however, we focus primarily on the restrictions placed on $\nu_{jt}$ and $\epsilon_{ijt}$ directly --- rather than the implications of these assumptions for the potential outcomes --- since this simplifies exposition and matches how these assumptions are stated in the literature.

\citet{donald_inference_2007} made an important early contribution to the literature on inference with few clusters. Their approach assumes that the cluster level-shocks $\nu_{jt}$ are mean-zero Gaussian, homoskedastic with respect to cluster and treatment status, and independent of other unit-and-time specific shocks. They also assume the number of units per cluster is large ($n_j \to \infty$ for all $j$). They then show that one can obtain valid inference by using critical values from a $t$-distribution with $J-2$ degrees of freedom, where $J$ is the total number of clusters. A nice feature of this approach is that it allows for valid inference when both the number of treated and untreated clusters is small. The disadvantage is the strong parametric assumption of homoskedastic Gaussian errors, which will often be hard to justify in practice.

\citet{Conley2011} introduce an alternative approach for inference that is able to relax the strong assumption of Gaussian errors in settings where there are many control clusters ($J_0$ large) but few treated clusters ($J_1$ small). This may be reasonable if the author has data from, say, 3 treated states and 47 untreated states. The key idea in \citet{Conley2011} is that if we assume that the errors in treated states come from the same distribution as in control states, then we can learn the distribution of errors from the large number of control states and use that to construct standard errors. A key advantage of this approach is that the distribution of errors is not assumed to be Gaussian, but rather is learned from the data. Nevertheless, the assumption that all treated groups have the same distribution of errors is still strong, and will often be violated if either there is heterogeneity in treatment effects or in cluster sizes. \citet{Ferman2018} extend the approach of \citet{Conley2011} to allow for heteroskedasticity caused by heterogeneity in group sizes or other observable characteristics, but must still restrict heterogeneity based on unobserved characteristics (e.g. unobserved treatment effect heterogeneity). 

\citet{Hagemann2020} provides an alternative permutation-based approach that avoids the need to directly estimate the heteroskedasticity. The key insight of \citet{Hagemann2020} is that if we place a bound on the maximal relative heterogeneity across clusters, then we can bound the probability of type I error from a permutation approach. He also shows how one can use this measure of relative heterogeneity to do sensitivity analysis. Like the other proposals above, though, \citet{Hagemann2020}'s approach must also place some strong restrictions on certain types of heterogeneity. In particular, his approach essentially requires that, as cluster size grows large, any single untreated cluster could be used to infer the counterfactual trend for the treated group, and thus his approach rules out cluster-specific heterogeneity in trends in untreated potential outcomes.

Another popular solution with few clusters is the cluster wild bootstrap. In an influential paper, \citet{Cameron2008} presented simulation evidence that the cluster wild bootstrap procedure can work well in settings with as few as five clusters. More recently, however, \citet{Canay2021} provided a formal analysis about the conditions under which the cluster wild bootstrap is asymptotically valid in settings with a few large clusters.  Importantly, \citet{Canay2021} show that the reliability of these bootstrap procedures depends on imposing certain homogeneity conditions on treatment effects, as well as the type of bootstrap weights one uses and the estimation method adopted (e.g., restricted vs. unrestricted OLS). These restrictions are commonly violated when one uses TWFE regressions with cluster-specific and time fixed effects like (\ref{eqn: static-TWFE-clusters}) or when treatment effects are allowed to be heterogeneous across clusters --- see Examples 2 and 3 in \citet{Canay2021}. Simulations have likewise shown that the cluster wild bootstrap may perform poorly in DiD settings with a small number of treated clusters \citep{mackinnon_wild_2018}. Thus, while the wild bootstrap may perform well in certain scenarios with a small number of clusters, it too requires strong homogeneity assumptions.

Finally, in settings with a large number of time periods, it may be feasible to conduct reliable inference with less stringent homogeneity assumptions about treatment effects. For instance, \citet{Canay2017}, \citet{Ibragimov2016}, \citet{Hagemann2021}, and \citet{chernozhukov_exact_2021} respectively propose permutation-based, $t$-test based, adjusted permutation-based, and conformal inference-based procedures that allow one to relax distributional assumptions about common shocks and accommodate richer forms of heterogeneity. The key restriction is that one is comfortable limiting the time-series dependence of the cluster-specific-shocks, and strengthening the parallel trends assumption to hold in many pre- and post-treatment time periods. These methods have been shown to be valid under asymptotics where the number of periods grows large. When in fact the number of time periods is small, as frequently occurs in DiD applications, one can still use some of these methods, but the underlying assumptions are stronger --- see, e.g., Remark 4.5 and Section 4.2 of \citet{Canay2017}. 

\paragraph{Alternative approaches.} We now briefly discuss two alternative approaches in settings with a small number of clusters. First, while all of the ``model-based'' papers above treat $\nu_{jt}$ as random, an alternative perspective would be to condition on the values of $\nu_{jt}$ and view the remaining uncertainty as coming only from the sampling of the individual units within clusters, constructing standard errors by clustering only at the unit level. This will generally produce a violation of parallel trends, but the violation may be relatively small if the cluster-specific shocks are small relative to the idiosyncratic variation. The violation of parallel trends could then be accounted for using the methods described in Section \ref{sec: pt violated}. To make this concrete, consider the setting of \citet{Card1994} that compares employment in NJ and PA after NJ raised its minimum wage. The aforementioned papers would consider NJ and PA as drawn from a super-population of treated and untreated states, where the state-level shocks are mean-zero, whereas the alternative approach would treat the two states as fixed and view any state-level shocks between NJ and PA as a violation of the parallel trends assumption. One could then explore the sensitivity of one's conclusions to the magnitude of this violation, potentially benchmarking it relative to the magnitude of the pre-treatment violations as discussed in Section \ref{subsec: sensitivity analysis}.

A second possibility is Fisher Randomization Tests (FRTs), otherwise known as permutation tests. The basic idea is to calculate some statistic of the data (e.g. the $t$-statistic of the DiD estimator), and then recompute this statistic under many permutations of the treatment assignment (at the cluster level). We then reject the null hypothesis of no effect if the test statistic using the original data is larger than 95\% of the draws of the test statistics under the permuted treatment assignment. Such tests have a long history in statistics, dating to \citet{fisher_design_1935}. If treatment is randomly assigned, then FRTs have exact finite-sample validity under the sharp null of no treatment effects for all units. The advantage of these tests is that they place no restrictions on the potential outcomes, and thus allow arbitrary heterogeneity in potential outcomes across clusters. On the other hand, the assumption of random treatment assignment may often be questionable in DiD settings. Moreover, the ``sharp'' null of no effects for all units may not be as economically interesting as the ``weak'' null of no average effects. Nevertheless, permutation tests may be a useful benchmark: if one cannot reject the null of no treatment effects \emph{even if treatment had been randomly assigned}, this suggests that there is not strong evidence of an effect in the data without other strong assumptions. In settings with staggered treatment timing, it may be more plausible to assume that the timing of when a unit gets treated is as good as random; see \citet{roth_efficient_2021} for efficient estimators and FRTs for this setting.

\paragraph{Recommendations.} In sum, recent research has made progress on the problem of conducting inference with relatively few clusters, but all of the available approaches require the researcher to impose some potentially strong additional assumptions. Most of the literature has focused on model-based approaches, which require the researcher to impose some homogeneity assumptions across clusters. Different homogeneity assumptions may be more reasonable in different contexts, and so we encourage researchers using these approaches to choose a method relying on a dimension of homogeneity that is most likely to hold (approximately) in their context. We also note that allowing for more heterogeneity may often come at the expense of obtaining tests with lower power. When none of these homogeneity assumptions is palatable, conditioning the inference on the cluster-level shocks and treating them as violations of parallel trends, accompanied by appropriate sensitivity analyses, may be an attractive alternative. Permutation-based methods also offer an intriguing alternative which requires no assumptions about homogeneity in potential outcomes, but requires stronger assumptions on the assignment of treatment and tests a potentially less interesting null hypothesis when the number of clusters is small.

\subsection{Design-based inference and the appropriate level of clustering\label{subsec: design based}}

The canonical approach to inference in DiD assumes that we have a large number of independently-drawn clusters sampled from an infinite super-population. In practice, however, there are two related conceptual difficulties with this framework. First, in many settings, it is unclear what the super-population of clusters is --- if the clusters in my sample are the 50 US states, should I view these as having been drawn from a super-population of possible states? Second, in many settings it is hard to determine what the appropriate level of clustering is --- if my data is on individuals who live in counties, which are themselves subsets of states, which is the appropriate level of clustering? 

To address these difficulties, it is often easier to consider a \emph{design-based} framework that views the units in the data as fixed (not necessarily sampled from a super-population) and the \emph{treatment assignment as stochastic}. This helps to address the difficulties described above, since we do not need to conceptualize the super-population, and the appropriate level of clustering is determined by the way that treatment is assigned. Design-based frameworks have a long history in statistics dating to \citet{neyman_application_1923}, and have received recent attention in econometrics \citep[e.g.][]{abadie_sampling-based_2020, AbadieEtAl(22)}. However, until recently, most of the results in the design-based literature has focused on settings where treatment probabilities are known or depend only on observable characteristics, and thus were not directly applicable to DiD.

Recent work by \citet{rambachan_design-based_2022} has extended this design-based view to settings like DiD, where treatment probabilities may differ in unknown ways across units. \citet{rambachan_design-based_2022} consider a setting similar to the canonical two-period model in Section \ref{sec: basic model}. However, following the design-based paradigm, they treat the units in the population (and their potential outcomes) as fixed rather than drawn from an infinite super-population. In this set-up, they show that the usual DiD estimator is unbiased for a finite-population analog to the ATT under a finite-population analog to the parallel trends assumption. In particular, let $\pi_i$ denote the probability that $D_i = 1$, and suppose that
$$ \sum_{i=1}^N \left(\pi_i - \frac{N_1}{N}\right) (Y_{i,2}(0) - Y_{i,1}(0)) = 0,$$
\noindent so that treatment probabilities are uncorrelated with trends in $Y(0)$ (a finite-population version of parallel trends). Then $E_D[\widehat{\tau}_2] = \tau_2^{Fin}$, where $\tau_2^{Fin} = E_D[ \frac{1}{N_1} \sum_{i:D_i=1}(Y_{i,2}(1) - Y_{i,2}(0))]$ is a finite-population analog to the ATT, i.e. the expected average treatment effect on the treated, where the expectation is taken over the stochastic distribution of which units are treated. 

\citet{rambachan_design-based_2022} show that from the design-based perspective, cluster-robust standard errors are valid (but potentially conservative) if the clustering is done at the level at which treatment is independently determined. Thus, for example, if the treatment is assigned independently at the unit-level,\footnote{Formally, if units are assigned independently before we condition on the number of treated units, $N_1$.} then we should cluster at the unit-level; by contrast, if treatment is determined independently across states, then we should cluster at the state level. This clear recommendation on the appropriate level of clustering contrasts with the more traditional model-based view that clustering should be done at the level at which the errors are correlated, which often makes it challenging to choose the appropriate level \citep{mackinnon_cluster-robust_2022}. These results also suggest that it may not actually be a problem if it is difficult to conceptualize a super-population from which our clusters are drawn; rather, the ``usual'' approach remains valid if there is no super-population and the uncertainty comes from stochastic assignment of treatment.\footnote{In some settings, the uncertainty may arise both from sampling and the stochastic assignment of treatment. \citet{AbadieEtAl(22)} study a model in which both treatment is stochastic and units are sampled from a larger population, and suggest that one should cluster among units if either their treatment assignments are correlated or the event that they are included in the sample is correlated. Although the \citet{AbadieEtAl(22)} results do not directly apply to DiD, we suspect that a similar heuristic would apply in DiD as well in light of the results in \citet{rambachan_design-based_2022} for the case where only treatment is viewed as stochastic. Formalizing this intuition strikes us as an interesting area for future research.}

\paragraph{Recommendations.} If it is difficult to conceptualize a super-population, fear not! Your DiD analysis can likely still be sensible from a finite-population perspective where we think of the treatment assignment as stochastic. Furthermore, if you are unsure about the appropriate level of clustering, a good rule of thumb (at least from the design-based perspective) is to cluster at the level at which treatment is independently assigned.

\section{Other topics and areas for future research \label{sec: other topics}}

In this section, we briefly touch on some other areas of interest in the DiD literature, and highlight some open areas for future research. 

\paragraph{Distributional treatment effects.} The DiD literature typically focuses on estimation of the ATT, but researchers may often be interested in the effect of a treatment on the entire distribution of an outcome. \citet{athey_identification_2006} propose the Changes-in-Changes model, which allows one to infer the full counterfactual distribution of $Y(0)$ for the treated group in DiD setups. The key assumption is that the mapping between quantiles of $Y(0)$ for the treated and comparison groups remains stable over time -- e.g., if the 30th percentile of the outcome for the treated group was the 70th percentile for the comparison group prior to treatment, this relationship would have been preserved in the second period if treatment had not occurred. \citet{bonhomme2011recovering} propose an alternative distributional DiD model based on a parallel trends assumption for the (log of the) characteristic function, which is motivated by a model of test scores. \citet{callaway_quantile_2019} propose a distributional DiD model based on a copula stability assumption. Finally, \citet{roth_when_2021} show that parallel trends holds for all functional forms under a ``parallel trends''-type assumption for the cumulative distribution of $Y(0)$, and this assumption also allows one to infer the full counterfactual distribution for the treated group. 

\paragraph{Quasi-random treatment timing.} In settings with staggered treatment timing, the generalized parallel trends assumption is often justified by arguing that the timing of treatment is random or quasi-random. \citet{roth_efficient_2021} show that if one is willing to assume treatment timing is as good as random, one can obtain more efficient estimates than using the staggered DiD methods discussed in Section \ref{subsec: staggered estimators}. This builds on earlier work by \citet{mckenzie_beyond_2012}, who highlighted that DiD is typically inefficient in an RCT where lagged outcomes are observed, as well as a large literature in statistics on efficient covariate adjustment in randomized experiments \citep[e.g.,][]{lin_agnostic_2013}. \citet{shaikh_randomization_2021} propose a method for observational settings where treatment timing is random conditional on fixed observable characteristics. We think that developing methods for observational settings where treatment timing is approximately random, possibly conditional on covariates and lagged outcomes, is an interesting area for further study in the years ahead. 

\paragraph{Sequential ignorability.} As discussed in Section \ref{subsec: related timing extensions}, an exciting new literature in DiD has begun to focus on settings where treatment can turn on and off and potential outcomes depend on the full path of treatments. A similar setting has been studied extensively in biostatistics, beginning with the pioneering work of \citet{Robins1986}. The key difference is that the biostatistics literature has focused on sequential random ignorability assumptions that impose that treatment in each period is random conditional on the path of covariates and realized outcomes, rather than parallel trends. We suspect that there may be economic settings where sequential ignorability may be preferable to parallel trends, e.g. when there is feedback between lagged outcomes and future treatment choices. Integrating these two literatures --- e.g., understanding in which economic settings is parallel trends preferable to sequential ignorability and vice versa --- is an interesting area for future research. An interesting step towards incorporating sequential ignorability in economic analyses is \citet{viviano_dynamic_2021}.

\paragraph{Spillover effects.} The vast majority of the DiD literature imposes the SUTVA assumption, which rules out spillover effects. However, spillover effects may be important in many economic applications, such as when policy in one area affects neighboring areas, or when individuals are connected in a network. \citet{butts_difference--differences_2021} provides some initial work in this direction by extending the framework of \citet{callaway_difference--differences_2020} to allow for local spatial spillovers. \citet{Huber2021} also consider extensions to allow for spillover effects. We suspect that in the coming years, we will see more work on DiD with spillovers. 

\paragraph{Conditional treatment effects.} The DiD literature has placed a lot of emphasis on learning about the ATT's of different groups. However, in many situations, it may also be desirable to better understand how these ATT's vary between subpopulations defined by covariate values. For instance, how does the average treatment effect of a training program on earnings vary according to the age of its participants? \citet{abadie_semiparametric_2005} provides re-weighting methods to tackle these types of questions using linear approximations. However, recent research has shown that data-adaptive/machine-learning procedures can be used to more flexibly estimate treatment effect heterogeneity in the context of RCTs or cross-sectional observational studies with unconfoundedness \citep[e.g.,][]{Lee2017, Wager2018,Chernozhukov2020}. Whether such tools can be adapted to estimate treatment effect heterogeneity in DiD setups is a promising area for future research.

\paragraph{Triple differences.} A common variant on DiD is triple-differences (DDD), which compares the DiD estimate for a demographic group expected to be affected by the treatment to a DiD for a second demographic group expected not to be affected (or effected less). For example, \citet{gruber_incidence_1994} studies the impacts of mandated maternity leave policies using a DDD design that compares the evolution of wages between treated/untreated states, before/after the law passed, and \textit{between} married women age 20-40 (who are expected to be affected) and other workers. DDD has received much less attention in the recent literature than standard DiD. We note, however, that DDD can often be cast as a DiD with a transformed outcome. For example, if we defined the state-level outcome $\tilde{Y}$ as the \textit{difference} in wages between women age 20-40 and other workers, then \citet{gruber_incidence_1994}'s DDD analysis would be equivalent to a DiD analysis using $\tilde{Y}$ as the outcome instead of wages. Nevertheless, we think that providing a more formal analysis of DDD along with practical recommendations for applied researchers would be a useful direction for future research. 

\paragraph{Connections to other panel data methods.} DiD is of course one of many possible panel data methods. One of the most prominent alternatives is the synthetic control (SC) method, pioneered by \citet{abadie_synthetic_2010}. Much of the DiD and SC literatures have evolved separately, using different data-generating processes as the baseline \citep{abadie_using_2021}. Recent work has begun to try to combine insights from the two literatures \citep[e.g.,][]{arkhangelsky_synthetic_2021, ben-michael_augmented_2021,Ben-Michael2021, doudchenko_balancing_2016}. We think that exploring further connections between the literatures --- and in particular, providing clear guidance for practitioners on when one we should expect one method to perform better than the other, or whether one should consider a hybrid of the two --- is an interesting direction for future research. 

\section{Conclusion}

This paper synthesizes the recent literature on DiD. Some key themes are that researchers should be clear about the comparison group used for identification, match the estimation and inference methods to the identifying assumptions, and explore robustness to possible violations of those assumptions. We emphasize that context-specific knowledge will often be needed to choose the right identifying assumptions and accompanying methods. We are hopeful that these recent developments will help to make DiD analyses more transparent and credible in the years to come.

\begin{table}[!ht]
\caption{A Checklist for DiD Practitioners\label{tbl: checklist}}
\begin{itemize}

\item \textbf{Is everyone treated at the same time?} 

If yes, and panel is balanced, estimation with TWFE specifications such as (\ref{eqn: static-TWFE-multiple-periods}) or (\ref{eqn:dynamic-twfe-multiple-periods}) yield easily interpretable estimates.

If no, consider using a ``heterogeneity-robust'' estimator for staggered treatment timing as described in Section \ref{sec:timing}. The appropriate estimator will depend on whether treatment turns on/off and which parallel trends assumption you're willing to impose. Use TWFE only if you're confident in treatment effect homogeneity. 

\item \textbf{Are you sure about the validity of the parallel trends assumption?}

If yes, explain why, including a justification for your choice of functional form. If the justification is (quasi-)random treatment timing, consider using a more efficient estimator as discussed in Section \ref{sec: other topics}.

If no, consider the following steps:

\begin{enumerate}
    \item 
    If parallel trends would be more plausible conditional on covariates, consider a method that conditions on covariates, as described in Section \ref{subsec: conditional pt}.
    
    \item
    Assess the plausibility of the parallel trends assumption by constructing an event-study plot. If there is a common treatment date and you're using an unconditional parallel trends assumption, plot the coefficients from a specification like (\ref{eqn: event-study spec}). If not, then see Section \ref{subsec: pre-trends testing} for recommendations on event-plot construction.

    \item 
    Accompany the event-study plot with diagnostics of the power of the pre-test against relevant alternatives and/or non-inferiority tests, as described in Section \ref{subsec: pretesting better}. 
    
    \item 
    Report formal sensitivity analyses that describe the robustness of the conclusions to potential violations of parallel trends, as described in Section \ref{subsec: sensitivity analysis}.

\end{enumerate}

\item \textbf{Do you have a large number of treated and untreated clusters sampled from a super-population?} 

If yes, then use cluster-robust methods at the cluster level. A good rule of thumb is to cluster at the level at which treatment is independently assigned (e.g. at the state level when policy is determined at the state level); see Section \ref{subsec: design based}.

If you have a small number of treated clusters, consider using one of the alternative inference methods described in Section \ref{subsec: model-based-inference}.

If you can't imagine the super-population, consider a design-based justification for inference instead, as discussed in Section \ref{subsec: design based}. 
\end{itemize}
    \label{tab:my_label}
\end{table}

\begin{table}[]
\caption{Statistical Packages for Recent DiD Methods}
\label{tbl:packages}
\centering
\par
\begin{adjustbox}{ max width=1\linewidth }
\begin{threeparttable}
\begin{tabular}{lll}\toprule
\noalign{\vskip 2mm} \multicolumn{3}{c}{\textbf{Heterogeneity Robust Estimators for Staggered Treatment Timing}}                                                                                                  \\\midrule
\uline{Package}                          & \uline{Software}       & \uline{Description}                                                                                                                            \\
\noalign{\vskip 2mm} did,   csdid                     & R, Stata       & Implements \citet{callaway_difference--differences_2020}                                                                                             \\
\noalign{\vskip 1.5mm} did2s                            & R, Stata       & Implements \citet{gardner_two-stage_2021},   \citet{borusyak_revisiting_2021}, \cite{sun_estimating_2020},     \\
& &  ~~~~~\citet{callaway_difference--differences_2020},   \citet{roth_efficient_2021} \\
\noalign{\vskip 1.5mm}didimputation,   did\_imputation \phantom{abc} & R, Stata \phantom{abc}      & Implements  \citet{borusyak_revisiting_2021}                                                                                    \\
\noalign{\vskip 1.5mm}DIDmultiplegt,   did\_multiplegt & R, Stata       & Implements \citet{de_chaisemartin_two-way_2020}                                                                           \\
\noalign{\vskip 1.5mm}eventstudyinteract               & Stata          & Implements \cite{sun_estimating_2020}                                                                                                  \\
\noalign{\vskip 1.5mm}flexpaneldid                     & Stata          & Implements \citet{dettmann_flexpaneldid_2020},   based on \citet{heckman_characterizing_1998}                                                                            \\
\noalign{\vskip 1.5mm}fixest                           & R              & Implements \cite{sun_estimating_2020}                                                                                                    \\
\noalign{\vskip 1.5mm}stackedev                        & Stata          & Implements stacking approach in   \citet{cengiz_effect_2019}                                                                                 \\
\noalign{\vskip 1.5mm}staggered                        & R              & Implements \citet{roth_efficient_2021}, \citet{callaway_difference--differences_2020},\\
& &  ~~~~~and \cite{sun_estimating_2020}  \\
\noalign{\vskip 1.5mm}xtevent                          & Stata          & Implements \citet{freyaldenhoven_pre-event_2019}                                                                                               \\\midrule

\noalign{\vskip 2mm} \multicolumn{3}{c}{\textbf{DiD with Covariates}}                                                                                                                                           \\\midrule
{\uline{Package}   }                    & {\uline{Software} } & {\uline{Description} }                                                                                                                      \\
\noalign{\vskip 2mm}DRDID, drdid                     & R, Stata       & Implements \citet{santanna_doubly_2020}                                                                                            \\\midrule
\noalign{\vskip 2mm} \multicolumn{3}{c}{\textbf{Diagnostics for TWFE   with Staggered Timing}}                                                                                                                  \\\midrule
\uline{Package}                                  & \uline{Software}       & \uline{Description}                                                                                                                             \\
\noalign{\vskip 1.5mm}bacondecomp, ddtiming                      & R, Stata       & Diagnostics from \citet{goodman-bacon_difference--differences_2018}                                                                                           \\
\noalign{\vskip 2mm}TwoWayFEWeights                  & R, Stata       & Diagnostics from \citet{de_chaisemartin_two-way_2020}                                                                            \\\midrule

\noalign{\vskip 2mm} \multicolumn{3}{c}{\textbf{Diagnostic /   Sensitivity for Violations of Parallel Trends}}                                                                                                  \\\midrule
\uline{Package}                             & \uline{Software}       & \uline{Description}                                                                                                                             \\
\noalign{\vskip 1.5mm}honestDiD                        & R, Stata       & Implements \citet{rambachan_honest_2021}                                                                                                   \\
\noalign{\vskip 1.5mm}pretrends                        & R              & Diagnostics from \citet{roth_pre-test_2021}          \\ \bottomrule
\end{tabular}
\begin{tablenotes}[para,flushleft]
\footnotesize{
Note:  This table lists R and Stata packages for recent DiD methods, and is based on Asjad Naqvi's repository at \href{https://asjadnaqvi.github.io/DiD/}{https://asjadnaqvi.github.io/DiD/}. Several of the packages listed under ``Heterogeneity Robust Estimators'' also accommodate covariates.}
\end{tablenotes}
\label{tab:addlabel}
\end{threeparttable}
\end{adjustbox}
\end{table}

\clearpage
\setlength\bibsep{0pt}
\bibliography{bibliography}

@Preamble{ " \newcommand{\noop}[1]{} " }

@article{abadie_robust_2022,
	title = {Robust {Post}-{Matching} {Inference}},
	volume = {117},
	issn = {0162-1459},
	url = {https://doi.org/10.1080/01621459.2020.1840383},
	doi = {10.1080/01621459.2020.1840383},
	number = {538},
	urldate = {2022-12-26},
	journal = {Journal of the American Statistical Association},
	author = {Abadie, Alberto and Spiess, Jann},
	month = apr,
	year = {2022},
	note = {Publisher: Taylor \& Francis
\_eprint: https://doi.org/10.1080/01621459.2020.1840383},
	pages = {983--995},
}

@misc{rambachan_design-based_2022,
	title = {Design-{Based} {Uncertainty} for {Quasi}-{Experiments}},
	url = {http://arxiv.org/abs/2008.00602},
	doi = {10.48550/arXiv.2008.00602},
	urldate = {2022-12-26},
	publisher = {arXiv},
	author = {Rambachan, Ashesh and Roth, Jonathan},
	month = nov,
	year = {2022},
	note = {arXiv:2008.00602 [econ, stat]},
}

@misc{de_chaisemartin_difference--differences_2022,
	title = {Difference-in-{Differences} {Estimators} of {Intertemporal} {Treatment} {Effects}},
	url = {http://arxiv.org/abs/2007.04267},
	doi = {10.48550/arXiv.2007.04267},
	urldate = {2022-12-26},
	publisher = {arXiv},
	author = {de Chaisemartin, Clément and D'Haultfoeuille, Xavier},
	month = mar,
	year = {2022},
	note = {arXiv:2007.04267 [econ]},
}

@techreport{schmidheiny_event_2020,
	address = {Rochester, NY},
	type = {{SSRN} {Scholarly} {Paper}},
	title = {On {Event} {Studies} and {Distributed}-{Lags} in {Two}-{Way} {Fixed} {Effects} {Models}: {Identification}, {Equivalence}, and {Generalization}},
	shorttitle = {On {Event} {Studies} and {Distributed}-{Lags} in {Two}-{Way} {Fixed} {Effects} {Models}},
	url = {https://papers.ssrn.com/abstract=3571164},
	language = {en},
	number = {3571164},
	urldate = {2022-06-08},
	institution = {Social Science Research Network},
	author = {Schmidheiny, Kurt and Siegloch, Sebastian},
	year = {2020},
	doi = {10.2139/ssrn.3571164},
}

@misc{viviano_dynamic_2021,
	title = {Dynamic covariate balancing: estimating treatment effects over time},
	shorttitle = {Dynamic covariate balancing},
	url = {http://arxiv.org/abs/2103.01280},
	doi = {10.48550/arXiv.2103.01280},
	urldate = {2022-12-23},
	publisher = {arXiv},
	author = {Viviano, Davide and Bradic, Jelena},
	month = jun,
	year = {2021},
	note = {arXiv:2103.01280 [econ, math, stat]},
}

@misc{caetano_difference_2022,
	title = {Difference in {Differences} with {Time}-{Varying} {Covariates}},
	url = {http://arxiv.org/abs/2202.02903},
	doi = {10.48550/arXiv.2202.02903},
	urldate = {2022-12-22},
	publisher = {arXiv},
	author = {Caetano, Carolina and Callaway, Brantly and Payne, Stroud and Rodrigues, Hugo Sant'Anna},
	month = feb,
	year = {2022},
	note = {arXiv:2202.02903 [econ]},
}

@article{bojinov_panel_2020,
	title = {Panel experiments and dynamic causal effects: A finite population perspective},
	author = {Bojinov, Iavor and Rambachan, Ashesh and Shephard, Neil},
	year = {2021},
    volume = {12},
    number = {4},
    pages = {1171--1196},
    journal = {Quantitative Economics}
}

@article{mackinnon_wild_2018,
	title = {The wild bootstrap for few (treated) clusters},
	volume = {21},
	issn = {1368-423X},
	url = {https://onlinelibrary.wiley.com/doi/abs/10.1111/ectj.12107},
	doi = {10.1111/ectj.12107},
	language = {en},
	number = {2},
	urldate = {2022-01-12},
	journal = {The Econometrics Journal},
	author = {MacKinnon, James G. and Webb, Matthew D.},
	year = {2018},
	pages = {114--135},
}

@article{chernozhukov_exact_2021,
	title = {An Exact and Robust Conformal Inference Method for Counterfactual and Synthetic Controls},
	journal = {Journal of the American Statistical Association},
	author = {Chernozhukov, Victor and Wüthrich, Kaspar and Zhu, Yinchu},
	year = {2021},
	volume = {116},
	number = {536},
	pages = {1849--1864 }
}

@article{gruber_incidence_1994,
	title = {The {Incidence} of {Mandated} {Maternity} {Benefits}},
	volume = {84},
	number = {3},
	urldate = {2022-01-12},
	journal = {The American Economic Review},
	author = {Gruber, Jonathan},
	year = {1994},
	pages = {622--641},
}

@article{callaway_difference--differences_2020,
	title = {Difference-in-{Differences} with multiple time periods},
	doi = {https://doi.org/10.1016/j.jeconom.2020.12.001},
	journal = {Journal of Econometrics},
	author = {Callaway, Brantly and Sant’Anna, Pedro H. C.},
	year = {2021},
	volume = {225},
	number = {2},
	pages = {200--230}
}

@article{de_chaisemartin_two-way_2020,
	title = {Two-{Way} {Fixed} {Effects} {Estimators} with {Heterogeneous} {Treatment} {Effects}},
	volume = {110},
	issn = {0002-8282},
	url = {https://www.aeaweb.org/articles?id=10.1257%2Faer.20181169},
	doi = {10.1257/aer.20181169},
	language = {en},
	number = {9},
	urldate = {2020-09-24},
	journal = {American Economic Review},
	author = {de Chaisemartin, Clément and D'Haultfœuille, Xavier},
	year = {2020},
	pages = {2964--2996},
}

@techreport{dettmann_flexpaneldid_2020,
	address = {Rochester, NY},
	type = {{SSRN} {Scholarly} {Paper}},
	title = {Flexpaneldid: {A} {Stata} {Toolbox} for {Causal} {Analysis} with {Varying} {Treatment} {Time} and {Duration}},
	shorttitle = {Flexpaneldid},
	url = {https://papers.ssrn.com/abstract=3692458},
	language = {en},
	number = {ID 3692458},
	urldate = {2021-12-30},
	institution = {Social Science Research Network},
	author = {Dettmann, Eva},
	year = {2020},
	doi = {10.2139/ssrn.3692458},
}

@article{baker_how_2021,
	title = {How much should we trust staggered difference-in-differences estimates?},
	journal = {Journal of Financial Economics},
	author = {Baker, Andrew and Larcker, David F. and Wang, Charles C. Y.},
	year = {2022},
	volume = {144},
	number = {2},
	pages = {370--395}
}

@article{Marcus2021,
author = {Marcus, Michelle and Sant'Anna, Pedro H. C.},
journal = {Journal of the Association of Environmental and Resource Economists},
number = {2},
pages = {235--275},
title = {{The role of parallel trends in event study settings : An application to environmental economics}},
volume = {8},
year = {2021}
}

@article{Wooldridge2021a,
author = {Wooldridge, Jeffrey M},
journal = {Working Paper},
pages = {1--89},
title = {{Two-Way Fixed Effects, the Two-Way Mundlak Regression, and Difference-in-Differences Estimators}},
year = {2021}
}

@article{butts_difference--differences_2021,
	title = {Difference-in-{Differences} {Estimation} with {Spatial} {Spillovers}},
	url = {http://arxiv.org/abs/2105.03737},
	urldate = {2021-12-29},
	journal = {arXiv:2105.03737 [econ]},
	author = {Butts, Kyle},
	year = {2021},
}

@article{Chang2020,
author = {Chang, Neng-Chieh},
doi = {10.1093/ectj/utaa001},
journal = {Econometrics Journal},
pages = {177--191},
title = {{Double/debiased machine learning for difference-in-differences}},
volume = {23},
year = {2020}
}

@article{Khan2010,
author = {Khan, Shakeeb and Tamer, Elie},
doi = {10.3982/ECTA7372},
issn = {0012-9682},
journal = {Econometrica},
number = {6},
pages = {2021--2042},
title = {{Irregular Identification, Support Conditions, and Inverse Weight Estimation}},
volume = {78},
year = {2010}
}

@article{Hagemann2021,
archivePrefix = {arXiv},
arxivId = {1907.01049},
author = {Hagemann, Andreas},
eprint = {1907.01049},
journal = {arXiv:1907.01049 [econ.EM]},
title = {{Permutation inference with a finite number of heterogeneous clusters}},
year = {2021}
}

@article{Meyer1995_jbes,
author = {Meyer, Bruce D.},
journal = {Journal of Business {\&} Economic Statistics},
number = {2},
pages = {151--161},
title = {{Natural and Quasi-Experiments in Economics}},
volume = {13},
year = {1995}
}

@article{Chernozhukov2020,
archivePrefix = {arXiv},
arxivId = {1712.04802},
author = {Chernozhukov, Victor and Demirer, Mert and Duflo, Esther and Fern{\'{a}}ndez-Val, Iv{\'{a}}n},
doi = {10.1920/wp.cem.2017.6117},
eprint = {1712.04802},
issn = {0898-2937},
journal = {arXiv: 1712.04802},
pages = {1--52},
title = {{Generic Machine Learning Inference on Heterogeneous Treatment Effects in Randomized Experiments}},
year = {2020}
}

@article{Heckman1997,
author = {Heckman, James J. and Ichimura, Hidehiko and Todd, Petra},
journal = {The Review of Economic Studies},
number = {4},
pages = {605--654},
title = {{Matching as an econometric evaluation estimator: Evidence from evaluating a job training programme}},
volume = {64},
year = {1997}
}

@article{Wager2018,
archivePrefix = {arXiv},
arxivId = {1510.04342},
author = {Wager, Stefan and Athey, Susan},
doi = {10.1080/01621459.2017.1319839},
eprint = {1510.04342},
issn = {1537274X},
journal = {Journal of the American Statistical Association},
number = {523},
pages = {1228--1242},
title = {{Estimation and Inference of Heterogeneous Treatment Effects using Random Forests}},
volume = {113},
year = {2018}
}

@article{Card1994,
author = {Card, David and Krueger, Alan B},
isbn = {00028282},
issn = {0002-8282},
journal = {American Economic Review},
number = {4},
pages = {772--793},
title = {{Minimum Wages and Employment: A Case Study of the Fast-Food Industry in New Jersey and Pennsylvania}},
volume = {84},
year = {1994}
}

@article{Lee2017,
archivePrefix = {arXiv},
arxivId = {1601.02801},
author = {Lee, Sokbae and Okui, Ryo and Whang, Yoon-Jae Jae},
doi = {10.1002/jae.2574},
eprint = {1601.02801},
issn = {10991255},
journal = {Journal of Applied Econometrics},
number = {7},
pages = {1207--1225},
title = {{Doubly robust uniform confidence band for the conditional average treatment effect function}},
volume = {32},
year = {2017}
}

@article{neyman_application_1923,
	title = {On the {Application} of {Probability} {Theory} to {Agricultural} {Experiments}. {Essay} on {Principles}. {Section} 9.},
	volume = {5},
	issn = {0883-4237},
	url = {https://www.jstor.org/stable/2245382},
	number = {4},
	urldate = {2020-04-21},
	journal = {Statistical Science},
	author = {Neyman, Jerzy},
	year = {1923},
	pages = {465--472}
}

@book{fisher_design_1935,
	address = {Oxford, England},
	series = {The design of experiments},
	title = {The design of experiments},
	publisher = {Oliver \& Boyd},
	author = {Fisher, R. A.},
	year = {1935},
	note = {Pages: xi, 251},
}

@article{Hagemann2020,
archivePrefix = {arXiv},
arxivId = {2010.04076},
author = {Hagemann, Andreas},
eprint = {2010.04076},
journal = {arXiv:2010.04076 [econ.EM]},
pages = {1--23},
title = {{Inference with a single treated cluster}},
year = {2020}
}

@article{lin_agnostic_2013,
	title = {Agnostic notes on regression adjustments to experimental data: {Reexamining} {Freedman}’s critique},
	volume = {7},
	issn = {1932-6157, 1941-7330},
	shorttitle = {Agnostic notes on regression adjustments to experimental data},
	url = {https://projecteuclid.org/euclid.aoas/1365527200},
	doi = {10.1214/12-AOAS583},
	language = {EN},
	number = {1},
	urldate = {2020-06-12},
	journal = {Annals of Applied Statistics},
	author = {Lin, Winston},
	year = {2013},
	mrnumber = {MR3086420},
	zmnumber = {06171273},
	pages = {295--318},
}

@article{shaikh_randomization_2021,
	title = {Randomization Tests in Observational Studies With Staggered Adoption of Treatment},
	journal = {Journal of the American Statistical Association},
	author = {Shaikh, Azeem and Toulis, Panos},
	year = {2021},
	volume = {116},
	number = {536},
	pages = {1835--1848}
}

@article{mckenzie_beyond_2012,
	title = {Beyond baseline and follow-up: {The} case for more {T} in experiments},
	volume = {99},
	issn = {0304-3878},
	shorttitle = {Beyond baseline and follow-up},
	url = {https://econpapers.repec.org/article/eeedeveco/v_3a99_3ay_3a2012_3ai_3a2_3ap_3a210-221.htm},
	number = {2},
	urldate = {2020-06-25},
	journal = {Journal of Development Economics},
	author = {McKenzie, David},
	year = {2012},
	pages = {210--221},
}

@article{roth_efficient_2021,
	title = {Efficient {Estimation} for {Staggered} {Rollout} {Designs}},
	url = {http://arxiv.org/abs/2102.01291},
	urldate = {2021-11-03},
	journal = {arXiv:2102.01291 [econ, math, stat]},
	author = {Roth, Jonathan and Sant'Anna, Pedro H. C.},
	year = {2021},
}

@article{callaway_quantile_2019,
	title = {Quantile treatment effects in difference in differences models with panel data},
	volume = {10},
	issn = {1759-7323},
	url = {http://qeconomics.org/ojs/index.php/qe/article/view/704},
	doi = {10.3982/QE935},
	language = {en},
	number = {4},
	urldate = {2020-04-23},
	journal = {Quantitative Economics},
	author = {Callaway, Brantly and Li, Tong},
	year = {2019},
	pages = {1579--1618},
}

@article{Conley2011,
author = {Conley, Timothy G. and Taber, Christopher R.},
doi = {10.1162/REST_a_00049},
issn = {0034-6535},
journal = {Review of Economics and Statistics},
number = {1},
pages = {113--125},
title = {{Inference with ``Difference in Differences'' with a Small Number of Policy Changes}},
volume = {93},
year = {2011}
}

@article{Canay2017,
author = {Canay, Ivan A. and Romano, Joseph P. and Shaikh, Azeem M.},
doi = {10.3982/ecta13081},
issn = {0012-9682},
journal = {Econometrica},
number = {3},
pages = {1013--1030},
title = {{Randomization Tests Under an Approximate Symmetry Assumption}},
volume = {85},
year = {2017}
}

@article{Ibragimov2016,
author = {Ibragimov, Rustam and M{\"{u}}ller, Ulrich K.},
doi = {10.1162/REST_a_00545},
issn = {15309142},
journal = {Review of Economics and Statistics},
number = {1},
pages = {83--96},
title = {{Inference with few heterogeneous clusters}},
volume = {98},
year = {2016}
}

@article{Canay2021,
author = {Canay, Ivan A. and Santos, Andres and Shaikh, Azeem M.},
doi = {10.1162/rest_a_00887},
issn = {15309142},
journal = {Review of Economics and Statistics},
number = {2},
pages = {346--363},
title = {{The wild bootstrap with a “small” number of “large” clusters}},
volume = {103},
year = {2021}
}

@article{Cameron2008,
author = {Cameron, A Colin and Gelbach, Jonah B and Miller, Douglas L},
journal = {Review of Economics and Statistics},
number = {3},
pages = {414--427},
title = {{Bootstrap-Based Improvements for Inference With Clustered Errors}},
volume = {90},
year = {2008}
}

@article{Ferman2018,
archivePrefix = {arXiv},
arxivId = {suresh govindarajan},
author = {Ferman, Bruno and Pinto, Cristine},
doi = {10.1162/rest_a_00759},
eprint = {suresh govindarajan},
isbn = {2007510134049},
issn = {0034-6535},
journal = {The Review of Economics and Statistics},
number = {3},
pages = {452--467},
pmid = {17411427},
title = {{Inference in Differences-in-Differences with Few Treated Groups and Heteroskedasticity}},
url = {https://www.mitpressjournals.org/doi/abs/10.1162/rest{\_}a{\_}00759},
volume = {101},
year = {2019}
}

@article{mackinnon_cluster-robust_2022,
	title = {Cluster-robust inference: A guide to empirical practice},
	journal = {Journal of Econometrics},
	author = {MacKinnon, James G. and Nielsen, Morten O. and Webb, Matthew D.},
	year = {2022},
	volume = {Forthcoming}
}

@article{abadie_sampling-based_2020,
	title = {Sampling-{Based} versus {Design}-{Based} {Uncertainty} in {Regression} {Analysis}},
	volume = {88},
	number = {1},
	journal = {Econometrica},
	author = {Abadie, Alberto and Athey, Susan and Imbens, Guido W. and Wooldridge, Jeffrey M.},
	year = {2020},
	pages = {265--296}
}

@article{roth_when_2021,
	title = {When {Is} {Parallel} {Trends} {Sensitive} to {Functional} {Form}?},
	journal = {Econometrica},
	author = {Roth, Jonathan and Sant'Anna, Pedro H. C.},
	year = {2022},
	volume = {Forthcoming}
}

@article{dette_difference--differences_2020,
	title = {Difference-in-{Differences} {Estimation} {Under} {Non}-{Parallel} {Trends}},
	journal = {Working Paper},
	author = {Dette, Holger and Schumann, Martin},
	year = {2020},
}

@article{roth_should_2018,
	title = {Should {We} {Adjust} for the {Test} for {Pre}-trends in {Difference}-in-{Difference} {Designs}?},
	url = {http://arxiv.org/abs/1804.01208},
	urldate = {2021-12-23},
	journal = {arXiv:1804.01208 [econ, math, stat]},
	author = {Roth, Jonathan},
	year = {2018},
}

@article{de_chaisemartin_two-way-survey_2021,
	title = {Two-way fixed effects and differences-in-differences with heterogeneous treatment effects: a survey},
	journal = {Econometrics Journal},
    volume = {Forthcoming},
	author = {de Chaisemartin, Clément and D'Haultfœuille, Xavier},
	year = {2022}
}

@article{callaway_difference--differences_2021,
	title = {Difference-in-{Differences} with a {Continuous} {Treatment}},
	url = {http://arxiv.org/abs/2107.02637},
	urldate = {2021-12-23},
	journal = {arXiv:2107.02637 [econ]},
	author = {Callaway, Brantly and Goodman-Bacon, Andrew and Sant'Anna, Pedro H. C.},
	year = {2021},
}

@techreport{de_chaisemartin_two-way_2021,
	address = {Rochester, NY},
	type = {{SSRN} {Scholarly} {Paper}},
	title = {Two-way {Fixed} {Effects} {Regressions} with {Several} {Treatments}},
	url = {https://papers.ssrn.com/abstract=3751060},
	language = {en},
	number = {ID 3751060},
	urldate = {2021-12-23},
	institution = {Social Science Research Network},
	author = {de Chaisemartin, Clément and D'Haultfœuille, Xavier},
	year = {2021},
	doi = {10.2139/ssrn.3751060},
}

@article{gardner_two-stage_2021,
	title = {Two-stage differences in differences},
	journal = {Working Paper},
	author = {Gardner, John},
	year={2021}
}

@article{borusyak_revisiting_2021,
	title = {Revisiting {Event} {Study} {Designs}: {Robust} and {Efficient} {Estimation}},
	shorttitle = {Revisiting {Event} {Study} {Designs}},
	url = {http://arxiv.org/abs/2108.12419},
	urldate = {2021-12-23},
	journal = {arXiv:2108.12419 [econ]},
	author = {Borusyak, Kirill and Jaravel, Xavier and Spiess, Jann},
	year = {2021},
}

@article{abadie_synthetic_2010,
	title = {Synthetic {Control} {Methods} for {Comparative} {Case} {Studies}: {Estimating} the {Effect} of {California}’s {Tobacco} {Control} {Program}},
	volume = {105},
	issn = {0162-1459, 1537-274X},
	shorttitle = {Synthetic {Control} {Methods} for {Comparative} {Case} {Studies}},
	url = {http://www.tandfonline.com/doi/abs/10.1198/jasa.2009.ap08746},
	doi = {10.1198/jasa.2009.ap08746},
	language = {en},
	number = {490},
	urldate = {2019-05-28},
	journal = {Journal of the American Statistical Association},
	author = {Abadie, Alberto and Diamond, Alexis and Hainmueller, Jens},
	month = jun,
	year = {2010},
	pages = {493--505}
}

@article{Ben-Michael2021,
author = {Ben-Michael, Eli and Feller, Avi and Rothstein, Jesse},
journal = {Journal of the Royal Statistical Society: Series B},
title = {{Synthetic controls with staggered adoption}},
volume = {84},
number = {2},
year = {2022},
pages = {351--381}
}

@article{bilinski_seeking_2018,
	title = {Seeking evidence of absence: {Reconsidering} tests of model assumptions},
	shorttitle = {Seeking evidence of absence},
	url = {http://arxiv.org/abs/1805.03273},
	journal = {arXiv:1805.03273 [stat]},
	author = {Bilinski, Alyssa and Hatfield, Laura A.},
	month = may,
	year = {2018}
}

@article{chabe-ferret_analysis_2015,
	title = {Analysis of the bias of {Matching} and {Difference}-in-{Difference} under alternative earnings and selection processes},
	volume = {185},
	issn = {0304-4076},
	url = {http://www.sciencedirect.com/science/article/pii/S0304407614002437},
	doi = {10.1016/j.jeconom.2014.09.013},
	number = {1},
	journal = {Journal of Econometrics},
	author = {Chabé-Ferret, Sylvain},
	year = {2015},
	pages = {110--123}
}

@article{daw_matching_2018,
	title = {Matching and {Regression} to the {Mean} in {Difference}‐in‐{Differences} {Analysis}},
	issn = {1475-6773},
	url = {https://onlinelibrary.wiley.com/doi/abs/10.1111/1475-6773.12993},
	doi = {10.1111/1475-6773.12993},
	language = {en},
	urldate = {2018-09-21},
	journal = {Health Services Research},
	author = {Daw, Jamie R. and Hatfield, Laura A.},
	year = {2018}
}

@article{athey_design-based_2018,
	title = {Design-based {Analysis} in {Difference}-{In}-{Differences} {Settings} with {Staggered} {Adoption}},
	journal = {Journal of Econometrics},
	author = {Athey, Susan and Imbens, Guido},
	volume = {226},
	number = {1},
	pages = {62--79},
	year = {2022}
}

@article{goodman-bacon_difference--differences_2018,
	title = {Difference-in-differences with variation in treatment timing},
	journal = {Journal of Econometrics},
	number = {2},
	volume = {225},
	pages = {254--277},
	author = {Goodman-Bacon, Andrew},
	year = {2021}
}

@article{athey_identification_2006,
	title = {Identification and {Inference} in {Nonlinear} {Difference}-in-{Differences} {Models}},
	volume = {74},
	issn = {0012-9682},
	url = {http://www.jstor.org/stable/3598807},
	number = {2},
	journal = {Econometrica},
	author = {Athey, Susan and Imbens, Guido W.},
	year = {2006},
	pages = {431--497}
}

@article{abadie_semiparametric_2005,
	title = {Semiparametric {Difference}-in-{Differences} {Estimators}},
	volume = {72},
	issn = {0034-6527},
	url = {http://www.jstor.org/stable/3700681},
	number = {1},
	journal = {The Review of Economic Studies},
	author = {Abadie, Alberto},
	year = {2005},
	pages = {1--19}
}

@article{bertrand_how_2004,
	title = {How {Much} {Should} {We} {Trust} {Differences}-{In}-{Differences} {Estimates}?},
	volume = {119},
	issn = {0033-5533},
	url = {https://academic-oup-com.ezp-prod1.hul.harvard.edu/qje/article/119/1/249/1876068},
	doi = {10.1162/003355304772839588},
	language = {en},
	number = {1},
	urldate = {2018-03-18},
	journal = {The Quarterly Journal of Economics},
	author = {Bertrand, Marianne and Duflo, Esther and Mullainathan, Sendhil},
	year = {2004},
	pages = {249--275}
}

@article{donald_inference_2007,
	title = {Inference with {Difference}-in-{Differences} and {Other} {Panel} {Data}},
	volume = {89},
	url = {https://ideas.repec.org/a/tpr/restat/v89y2007i2p221-233.html},
	language = {en},
	number = {2},
	urldate = {2018-03-18},
	journal = {The Review of Economics and Statistics},
	author = {Donald, Stephen G. and Lang, Kevin},
	year = {2007},
	pages = {221--233}
}

@techreport{borusyak_revisiting_2016,
	address = {Rochester, NY},
	type = {{SSRN} {Scholarly} {Paper}},
	title = {Revisiting {Event} {Study} {Designs}},
	url = {https://papers.ssrn.com/abstract=2826228},
	language = {en},
	number = {ID 2826228},
	urldate = {2018-03-15},
	institution = {Social Science Research Network},
	author = {Borusyak, Kirill and Jaravel, Xavier},
	year = {2018}
}

@article{Huber2021,
author = {Huber, Martin and Steinmayr, Andreas},
doi = {10.1080/07350015.2019.1668795},
journal = {Journal of Business {\&} Economic Statistics},
number = {2},
pages = {422--436},
title = {{A Framework for Separating Individual-Level Treatment Effects From Spillover Effects}},
volume = {39},
year = {2021}
}

@article{andrews_inference_2019,
	title = {Inference for {Linear} {Conditional} {Moment} {Inequalities}},
	author = {Andrews, Isaiah and Roth, Jonathan and Pakes, Ariel},
	year = {2022},
	journal = {Review of Economic Studies},
	volume = {Forthcoming}
}

@article{Ding2019,
author = {Ding, Peng and Li, Fan},
journal = {Political Analysis},
pages = {605----615},
title = {{A Bracketing Relationship between Difference-in-Differences and Lagged-Dependent-Variable Adjustment}},
volume = {27},
year = {2019}
}

@article{de_chaisemartin_fuzzy_2018,
	title = {Fuzzy {Differences}-in-{Differences}},
	volume = {85},
	issn = {0034-6527},
	url = {https://academic.oup.com/restud/article/85/2/999/4096388},
	doi = {10.1093/restud/rdx049},
	language = {en},
	number = {2},
	urldate = {2018-09-21},
	journal = {The Review of Economic Studies},
	author = {de Chaisemartin, Clément and D'Haultfœuille, Xavier},
	year = {2018},
	pages = {999--1028}
}

@article{manski_how_2017,
	title = {How {Do} {Right}-to-{Carry} {Laws} {Affect} {Crime} {Rates}? {Coping} with {Ambiguity} {Using} {Bounded}-{Variation} {Assumptions}},
	volume = {100},
	issn = {0034-6535},
	shorttitle = {How {Do} {Right}-to-{Carry} {Laws} {Affect} {Crime} {Rates}?},
	url = {https://doi.org/10.1162/REST_a_00689},
	doi = {10.1162/REST_a_00689},
	number = {2},
	urldate = {2019-07-19},
	journal = {The Review of Economics and Statistics},
	author = {Manski, Charles F. and Pepper, John V.},
	year = {2018},
	pages = {232--244}
}

@article{Rubin1974,
author = {Rubin, Donald B.},
journal = {Journal of Educational Psychology},
number = {5},
pages = {688--70},
title = {{Estimating causal effects of treatments in randomized and nonrandomized studies}},
volume = {66},
year = {1974}
}

@article{Wooldridge2003,
author = {Wooldridge, Jeffrey M},
journal = {American Economic Review P{\&}P},
number = {2},
pages = {133--138},
title = {{Cluster-Sample Methods in Applied Econometrics}},
volume = {93},
year = {2003}
}

@article{Robins1986,
author = {Robins, J. M.},
doi = {10.1016/0898-1221(87)90238-0},
issn = {00974943},
journal = {Mathematical Modelling},
pages = {1393--1512},
title = {{A New Approach To Causal Inference in Mortality Studies With a Sustained Exposure Period - Application To Control of the Healthy Worker Survivor Effect}},
volume = {7},
year = {1986}
}

@article{Abbring2003,
author = {Abbring, Jaap H. and van den Berg, Gerard J.},
journal = {Econometrica},
number = {5},
pages = {1491--1517},
title = {{The nonparametric identification of treatment effects in duration models}},
volume = {71},
year = {2003}
}

@article{liang_longitudinal_1986,
	title = {Longitudinal data analysis using generalized linear models},
	volume = {73},
	issn = {0006-3444},
	url = {https://academic.oup.com/biomet/article/73/1/13/246001},
	doi = {10.1093/biomet/73.1.13},
	language = {en},
	number = {1},
	urldate = {2020-03-10},
	journal = {Biometrika},
	author = {Liang, Kung-Yee and Zeger, Scott L.},
	year = {1986},
	pages = {13--22}
}

@article{arellano_practitioners_1987,
	title = {{Practitioners}’ {Corner}: {Computing} {Robust} {Standard} {Errors} for {Within}-groups {Estimators}},
	volume = {49},
	copyright = {© 1987 Blackwell Publishing Ltd},
	issn = {1468-0084},
	shorttitle = {{PRACTITIONERS}’ {CORNER}},
	url = {https://onlinelibrary.wiley.com/doi/abs/10.1111/j.1468-0084.1987.mp49004006.x},
	doi = {10.1111/j.1468-0084.1987.mp49004006.x},
	language = {en},
	number = {4},
	urldate = {2020-03-10},
	journal = {Oxford Bulletin of Economics and Statistics},
	author = {Arellano, M.},
	year = {1987},
	pages = {431--434}
}

@article{malani_interpreting_2015,
	title = {Interpreting pre-trends as anticipation: {Impact} on estimated treatment effects from tort reform},
	volume = {124},
	issn = {0047-2727},
	shorttitle = {Interpreting pre-trends as anticipation},
	url = {http://www.sciencedirect.com/science/article/pii/S0047272715000122},
	doi = {10.1016/j.jpubeco.2015.01.001},
	language = {en},
	urldate = {2019-10-25},
	journal = {Journal of Public Economics},
	author = {Malani, Anup and Reif, Julian},
	year = {2015},
	pages = {1--17}
}

@article{holland_statistics_1986,
	title = {Statistics and {Causal} {Inference}},
	volume = {81},
	issn = {0162-1459},
	url = {https://www.jstor.org/stable/2289064},
	doi = {10.2307/2289064},
	number = {396},
	urldate = {2020-04-07},
	journal = {Journal of the American Statistical Association},
	author = {Holland, Paul W.},
	year = {1986},
	pages = {945--960}
}

@article{strezhnev2018semiparametric,
  title={Semiparametric weighting estimators for
multi-period difference-in-differences designs},
  author={Strezhnev, Anton},
  year={2018},
  journal = {Working Paper}
}

@article{cengiz_effect_2019,
	title = {The {Effect} of {Minimum} {Wages} on {Low}-{Wage} {Jobs}},
	volume = {134},
	issn = {0033-5533},
	url = {https://academic.oup.com/qje/article/134/3/1405/5484905},
	doi = {10.1093/qje/qjz014},
	language = {en},
	number = {3},
	urldate = {2020-04-07},
	journal = {The Quarterly Journal of Economics},
	author = {Cengiz, Doruk and Dube, Arindrajit and Lindner, Attila and Zipperer, Ben},
	year = {2019},
	pages = {1405--1454}
}

@Article{bonhomme2011recovering,
  author    = {Bonhomme, St{\'e}phane and Sauder, Ulrich},
  journal   = {Review of Economics and Statistics},
  title     = {Recovering distributions in difference-in-differences models: A comparison of selective and comprehensive schooling},
  year      = {2011},
  number    = {2},
  pages     = {479--494},
  volume    = {93},
  publisher = {MIT Press},
}

@article{freyaldenhoven_pre-event_2019,
	title = {Pre-event {Trends} in the {Panel} {Event}-{Study} {Design}},
	volume = {109},
	issn = {0002-8282},
	url = {https://www.aeaweb.org/articles?id=10.1257/aer.20180609},
	doi = {10.1257/aer.20180609},
	language = {en},
	number = {9},
	urldate = {2021-07-16},
	journal = {American Economic Review},
	author = {Freyaldenhoven, Simon and Hansen, Christian and Shapiro, Jesse M.},
	year = {2019},
	pages = {3307--3338},
}

@article{kahn-lang_promise_2020,
	title = {The {Promise} and {Pitfalls} of {Differences}-in-{Differences}: {Reflections} on 16 and {Pregnant} and {Other} {Applications}},
	volume = {38},
	issn = {0735-0015},
	shorttitle = {The {Promise} and {Pitfalls} of {Differences}-in-{Differences}},
	url = {https://doi.org/10.1080/07350015.2018.1546591},
	doi = {10.1080/07350015.2018.1546591},
	number = {3},
	urldate = {2021-07-16},
	journal = {Journal of Business \& Economic Statistics},
	author = {Kahn-Lang, Ariella and Lang, Kevin},

	year = {2020},
	pages = {613--620},
}

@article{roth_pre-test_2021,
	title = {Pre-test with {Caution}: {Event}-study {Estimates} {After} {Testing} for {Parallel} {Trends}},
	year = {2022},
	author = {Roth, Jonathan},
	journal = {American Economic Review: Insights},
	volume = {4},
	number = {3},
	pages = {305--322}
}

@article{hasegawa_evaluating_2019,
	title = {Evaluating {Missouri}’s {Handgun} {Purchaser} {Law}: {A} {Bracketing} {Method} for {Addressing} {Concerns} {About} {History} {Interacting} with {Group}},
	volume = {30},
	issn = {1044-3983},
	number = {3},
	urldate = {2021-09-23},
	journal = {Epidemiology},
	author = {Hasegawa, Raiden B. and Webster, Daniel W. and Small, Dylan S.},
	year = {2019},
	pages = {371--379},
}

@article{ye_negative_2021,
	title = {A {Negative} {Correlation} {Strategy} for {Bracketing} in {Difference}-in-{Differences}},
	url = {http://arxiv.org/abs/2006.02423},
	urldate = {2021-09-24},
	journal = {arXiv:2006.02423 [econ, stat]},
	author = {Ye, Ting and Keele, Luke and Hasegawa, Raiden and Small, Dylan S.},
	year = {2021},
}

@article{keele_patterns_2019,
	title = {Patterns of {Effects} and {Sensitivity} {Analysis} for {Differences}-in-{Differences}},
	url = {http://arxiv.org/abs/1901.01869},
	urldate = {2021-09-27},
	journal = {arXiv:1901.01869 [stat]},
	author = {Keele, Luke J. and Small, Dylan S. and Hsu, Jesse Y. and Fogarty, Colin B.},
	month = feb,
	year = {2019},
	note = {arXiv: 1901.01869},
}

@article{liu_practical_2021,
	title = {A Practical Guide to Counterfactual Estimators for Causal Inference with Time-Series Cross-Sectional Data},
	volume = {Forthcoming},
    journal = {American Journal of Political Science},
	author = {Liu, Licheng and Wang, Ye and Xu, Yiqing},
	year = {2022}
}

@article{armstrong_optimal_2018,
	title = {Optimal {Inference} in a {Class} of {Regression} {Models}},
	volume = {86},
	issn = {1468-0262},
	url = {https://onlinelibrary.wiley.com/doi/abs/10.3982/ECTA14434},
	doi = {10.3982/ECTA14434},
	language = {en},
	number = {2},
	urldate = {2021-12-27},
	journal = {Econometrica},
	author = {Armstrong, Timothy B. and Kolesár, Michal},
	year = {2018},
	pages = {655--683},
}

@article{olea_simultaneous_2019,
	title = {Simultaneous confidence bands: {Theory}, implementation, and an application to {SVARs}},
	volume = {34},
	copyright = {© 2018 John Wiley \& Sons, Ltd.},
	issn = {1099-1255},
	shorttitle = {Simultaneous confidence bands},
	url = {https://onlinelibrary.wiley.com/doi/abs/10.1002/jae.2656},
	doi = {https://doi.org/10.1002/jae.2656},
	language = {en},
	number = {1},
	urldate = {2021-03-05},
	journal = {Journal of Applied Econometrics},
	author = {Olea, José Luis Montiel and Plagborg‐Møller, Mikkel},
	year = {2019},
	pages = {1--17},
}

@article{freyaldenhoven_visualization_2021,
	title = {Visualization, identification, and estimation in the linear panel event-study design.},
	journal = {Advances in Economics and Econometrics: Twelfth World Congress},
	year = {2021},
	volume = {Forthcoming},
	author = {Simon Freyaldenhoven and Christian Hansen and Jorge P{\'e}rez P{\'e}rez and Jesse M. Shapiro}
}

@article{rambachan_honest_2021,
	title = {A More Credible Approach to Parallel Trends },
	year = 2022,
	author = {Rambachan, Ashesh and Roth, Jonathan},
	journal = {Review of Economic Studies},
	volume = {Forthcoming}
}

@article{heckman_characterizing_1998,
	title = {Characterizing {Selection} {Bias} {Using} {Experimental} {Data}},
	volume = {66},
	issn = {0012-9682},
	url = {https://www.jstor.org/stable/2999630},
	doi = {10.2307/2999630},
	number = {5},
	urldate = {2021-12-23},
	journal = {Econometrica},
	author = {Heckman, James and Ichimura, Hidehiko and Smith, Jeffrey and Todd, Petra},
	year = {1998},
	pages = {1017--1098},
}

@article{sun_estimating_2020,
	title = {Estimating dynamic treatment effects in event studies with heterogeneous treatment effects},
	doi = {10.1016/j.jeconom.2020.09.006},
	language = {en},
	journal = {Journal of Econometrics},
	author = {Sun, Liyang and Abraham, Sarah},
	year = {2021},
	volume = {225},
	number = {2},
	pages = {175--199}
}

@article{santanna_doubly_2020,
	title = {Doubly robust difference-in-differences estimators},
	volume = {219},
	issn = {0304-4076},
	url = {https://www.sciencedirect.com/science/article/pii/S0304407620301901},
	doi = {10.1016/j.jeconom.2020.06.003},
	language = {en},
	number = {1},
	urldate = {2021-12-23},
	journal = {Journal of Econometrics},
	author = {Sant'Anna, Pedro H. C. and Zhao, Jun},
	year = {2020},
	pages = {101--122},
}

@article{zeldow_confounding_2021,
	title = {Confounding and regression adjustment in difference-in-differences studies},
	volume = {56},
	issn = {1475-6773},
	doi = {10.1111/1475-6773.13666},
	number = {5},
	journal = {Health Services Research},
	author = {Zeldow, Bret and Hatfield, Laura A.},
	year = {2021},
	pmid = {33978956},
	pmcid = {PMC8522571},
	pages = {932--941},
}

@article{ryan_well-balanced_2018,
	title = {Well-{Balanced} or too {Matchy}–{Matchy}? {The} {Controversy} over {Matching} in {Difference}-in-{Differences}},
	volume = {53},
	issn = {1475-6773},
	shorttitle = {Well-{Balanced} or too {Matchy}–{Matchy}?},
	url = {https://onlinelibrary.wiley.com/doi/abs/10.1111/1475-6773.13015},
	doi = {10.1111/1475-6773.13015},
	language = {en},
	number = {6},
	urldate = {2021-12-23},
	journal = {Health Services Research},
	author = {Ryan, Andrew M.},
	year = {2018},
	pages = {4106--4110},
}

@article{jakiela_simple_2021,
	title = {Simple {Diagnostics} for {Two}-{Way} {Fixed} {Effects}},
	url = {http://arxiv.org/abs/2103.13229},
	urldate = {2021-12-28},
	journal = {arXiv:2103.13229 [econ, q-fin]},
	author = {Jakiela, Pamela},
	month = mar,
	year = {2021},
}

@article{arkhangelsky_synthetic_2021,
	title = {Synthetic {Difference}-in-{Differences}},
	volume = {111},
	issn = {0002-8282},
	url = {https://www.aeaweb.org/articles?id=10.1257/aer.20190159},
	doi = {10.1257/aer.20190159},
	language = {en},
	number = {12},
	urldate = {2021-12-28},
	journal = {American Economic Review},
	author = {Arkhangelsky, Dmitry and Athey, Susan and Hirshberg, David A. and Imbens, Guido W. and Wager, Stefan},
	year = {2021},
	pages = {4088--4118},
}

@article{Abadie2006,
author = {Abadie, Alberto and Imbens, Guido W.},
journal = {Econometrica},
number = {1},
pages = {235--267},
title = {{Large sample properties of matching estimators for average treatment effects}},
volume = {74},
year = {2006}
}

@article{Abadie2012,
author = {Abadie, Alberto and Imbens, Guido W.},
doi = {10.1080/01621459.2012.682537},
issn = {0162-1459},
journal = {Journal of the American Statistical Association},
number = {498},
pages = {833--843},
title = {{A Martingale Representation for Matching Estimators}},
volume = {107},
year = {2012}
}

@article{Abadie2011,
author = {Abadie, Alberto and Imbens, Guido W.},
doi = {10.1198/jbes.2009.07333},
isbn = {978-90-481-8767-6},
issn = {0735-0015},
journal = {Journal of Business {\&} Economic Statistics},
number = {1},
pages = {1--11},
title = {{Bias-Corrected Matching Estimators for Average Treatment Effects}},
volume = {29},
year = {2011}
}

@article{Abadie2008,
author = {Abadie, Alberto and Imbens, Guido W.},
doi = {10.3982/ECTA6474},
issn = {0012-9682},
journal = {Econometrica},
number = {6},
pages = {1537--1557},
title = {{On the Failure of the Bootstrap for Matching Estimators}},
url = {http://doi.wiley.com/10.3982/ECTA6474},
volume = {76},
year = {2008}
}

@article{ben-michael_augmented_2021,
	title = {The {Augmented} {Synthetic} {Control} {Method}},
	volume = {116},
	issn = {0162-1459},
	url = {https://doi.org/10.1080/01621459.2021.1929245},
	doi = {10.1080/01621459.2021.1929245},
	number = {536},
	urldate = {2021-12-28},
	author = {Ben-Michael, Eli and Feller, Avi and Rothstein, Jesse},
	journal = {Journal of the American Statistical Association},
	year = {2021},
	pages = {1789--1803},
}

@article{abadie_using_2021,
	title = {Using {Synthetic} {Controls}: {Feasibility}, {Data} {Requirements}, and {Methodological} {Aspects}},
	volume = {59},
	issn = {0022-0515},
	shorttitle = {Using {Synthetic} {Controls}},
	url = {https://pubs.aeaweb.org/doi/10.1257/jel.20191450},
	doi = {10.1257/jel.20191450},
	language = {en},
	number = {2},
	urldate = {2021-12-28},
	journal = {Journal of Economic Literature},
	author = {Abadie, Alberto},
	month = jun,
	year = {2021},
	pages = {391--425},
}

@techreport{doudchenko_balancing_2016,
	type = {Working {Paper}},
	title = {Balancing, {Regression}, {Difference}-{In}-{Differences} and {Synthetic} {Control} {Methods}: {A} {Synthesis}},
	shorttitle = {Balancing, {Regression}, {Difference}-{In}-{Differences} and {Synthetic} {Control} {Methods}},
	url = {https://www.nber.org/papers/w22791},
	number = {22791},
	urldate = {2021-12-28},
	institution = {National Bureau of Economic Research},
	author = {Doudchenko, Nikolay and Imbens, Guido W.},
	year = {2016},
	doi = {10.3386/w22791},
}

\appendix

\section{Connecting model-based assumptions to potential outcomes\label{sec: model based appendix}}
This section formalizes connections between the model-based assumptions in Section \ref{subsec: model-based-inference} and the potential outcomes framework. We derive how the errors in the structural model (\ref{eqn: static-TWFE-ind-clusters}) map to primitives based on potential outcomes in the canonical model from Section \ref{sec: basic model}. Specifically, we show that under the set-up of Section \ref{sec: basic model}, Assumptions \ref{asm: parallel-trends-2-periods} and \ref{asm: no anticipation} imply that the canonical DiD estimator takes the form given in (\ref{eqn: hatbeta-structural-model}), where $\beta = \tau_2$ is the ATT at the cluster level, $\nu_{jt} = \nu_{jt,0} + D_j \nu_{jt,1}$ and $\epsilon_{ijt} = \epsilon_{ijt,0} + D_j \epsilon_{ijt,1}$, where\footnote{In what follows, we write $\expe{Y_{ijt}(0)|D_j=d}$ to denote the expectation where one first draws $j$ from the population with $D_j=d$ and then draws $Y_{ijt}(0)$ from that cluster.} \begin{align*}
&\epsilon_{ijt,0} = Y_{ijt}(0) - \expe{Y_{ijt}(0) | j(i) = j} &\\
&\epsilon_{ijt,1} = Y_{ijt}(1) - Y_{ijt}(0) - \expe{Y_{ijt}(1) - Y_{ijt}(0) | j(i) = j}& \\
&\nu_{jt,0} = \expe{Y_{ijt}(0) | j(i) = j} - \expe{Y_{ijt}(0) | D_j} &\\
&\nu_{jt,1} = \expe{Y_{ijt}(1) - Y_{ijt}(0)|j(i) =j} - \tau_t.
\end{align*}
\noindent Thus, in the canonical set-up, restrictions on $\nu_{jt}$ and $\epsilon_{ijt}$ can be viewed as restrictions on primitives that are functions of the potential outcomes. 

Adopt the notation and set-up in Section \ref{sec: basic model}, except now each unit $i$ belongs to a cluster $j$ and treatment is assigned at the cluster level $D_j$. We assume clusters are drawn $iid$ from a superpopulation of clusters and then units are drawn $iid$ within the sampled cluster. We write $J_d$ to denote the number of clusters with treatment $d$, and $n_j$ the number of units in cluster $j$. As in the main text, let $Y_{jt} = {n_j}^{-1} \sum_{i:j(i)=j} Y_{ijt}$ be the sample mean within cluster $j$. The canonical DiD estimator at the cluster level can then be written as: 
\begin{align*}
\widehat\tau &= \frac{1}{J_1} \sum_{j:D_j =1} (Y_{j2} - Y_{j1}) - \frac{1}{J_0} \sum_{i:D_j =0} (Y_{j2} - Y_{j1}) \\
&= \frac{1}{J_1} \sum_{j:D_j =1} \frac{1}{n_j} \sum_{i:j(i)=j} (Y_{ij2} - Y_{ij1}) - \frac{1}{J_0} \sum_{i:D_j =0}  \frac{1}{n_j} \sum_{i:j(i)=j} (Y_{ij2} - Y_{ij1}).
\end{align*}

\noindent Since the observed outcome is $Y(1)$ for treated units and $Y(0)$ for control units, under the no anticipation assumption it follows that
\begin{align*}
\widehat\tau &= \frac{1}{J_1} \sum_{j:D_j =1} \frac{1}{n_j} \sum_{i:j(i)=j} (Y_{ij2}(1) - Y_{ij1}(0)) - \frac{1}{J_0} \sum_{j:D_j =0} \frac{1}{n_j} \sum_{i:j(i)=j} (Y_{ij2}(0) - Y_{ij1}(0)),
\end{align*}
\noindent or equivalently,
\begin{align*}
\widehat\tau =& \frac{1}{J_1} \sum_{j:D_j =1} \frac{1}{n_j} \sum_{i:j(i)=j} (Y_{ij2}(1) - Y_{ij2}(0)) +  \\
& \frac{1}{J_1} \sum_{j:D_j =1} \frac{1}{n_j} \sum_{i:j(i)=j} (Y_{ij2}(0) - Y_{ij1}(0)) -\\ & \frac{1}{J_0} \sum_{j:D_j =0}  \frac{1}{n_j} \sum_{i:j(i)=j} (Y_{ij2}(0) - Y_{ij1}(0)).
\end{align*}

\noindent Adding and subtracting terms of the form $\expe{Y_{ijt} | j(i) = 1}$, we obtain
\begin{align*}
\widehat\tau =& \tau_{2} + \frac{1}{J_1} \sum_{j:D_j=1} (\expe{Y_{ij2}(1) - Y_{ij2}(0) | j(i) =j } - \tau_2) +\\&
\frac{1}{J_1} \sum_{j:D_j =1} \frac{1}{n_j} \sum_{i:j(i)=j} (Y_{ij2}(1) - Y_{ij2}(0)- \expe{Y_{ij2}(1) - Y_{ij2}(0) | j(i) =j } ) +  \\
& \frac{1}{J_1} \sum_{j:D_j =1} \frac{1}{n_j} \sum_{i:j(i)=j} (Y_{ij2}(0) - Y_{ij1}(0) - \expe{Y_{ij2}(0) - Y_{ij1}(0) | j(i) =j}) -\\ & \frac{1}{J_0} \sum_{j:D_j =0}  \frac{1}{n_j} \sum_{i:j(i)=j} (Y_{ij2}(0) - Y_{ij1}(0) - \expe{Y_{ij2}(0) - Y_{ij1}(0) | j(i) =j}) + \\
& \frac{1}{J_1} \sum_{j:D_j=1} \expe{Y_{ij2}(0) - Y_{ij1}(0) | j(i) =j} - \frac{1}{J_0} \sum_{j:D_j=0} \expe{Y_{ij2}(0) - Y_{ij1}(0) | j(i) =j}  ,
\end{align*}

\noindent where $\tau_2 = \expe{ {J_1}^{-1} \sum_{j:D_j=1} \expe{Y_{ij2}(1) - Y_{ij2}(0)  | j(i) = j} } = \expe{Y_{ij2}(1) - Y_{ij2}(0)| D_j = 1}$ is the ATT among treated clusters (weighting all clusters equally).

Now, we assume parallel trends at the cluster level, so that
$$ \expe{Y_{ij2}(0) - Y_{ij1}(0) | D_j = 1} - \expe{Y_{ij2}(0) - Y_{ij1}(0) | D_j =0}  = 0,$$
\noindent which implies that

\begin{align*}
\widehat\tau =& \tau_{2} + \frac{1}{J_1} \sum_{j:D_j=1} \underbrace{(\expe{Y_{ij2}(1) - Y_{ij2}(0) | j(i) =j } - \tau_2)}_{=\Delta\nu_{j,1}} +\\&
\frac{1}{J_1} \sum_{j:D_j =1} \frac{1}{n_j} \sum_{i:j(i)=j} \underbrace{(Y_{ij2}(1) - Y_{ij2}(0)-\expe{Y_{ij2}(1) - Y_{ij2}(0) | j(i) =j } )}_{=\Delta\epsilon_{ij,1}} +  \\
& \frac{1}{J_1} \sum_{j:D_j =1} \frac{1}{n_j} \sum_{i:j(i)=j} \underbrace{(Y_{ij2}(0) - Y_{ij1}(0) - \expe{Y_{ij2}(0) - Y_{ij1}(0) | j(i) =j})}_{=\Delta\epsilon_{ij,0}} -\\ & \frac{1}{J_0} \sum_{i:D_j =0}  \frac{1}{n_j} \sum_{i:j(i)=j} \underbrace{(Y_{ij2}(0) - Y_{ij1}(0) - \expe{Y_{ij2}(0) - Y_{ij1}(0) | j(i) =j})}_{=\Delta\epsilon_{ij,0}} + \\
& \frac{1}{J_1} \sum_{j:D_j=1} \underbrace{(\expe{Y_{ij2}(0) - Y_{ij1}(0) | j(i) =j} - \expe{Y_{ij2}(0) - Y_{ij1}(0) | D_j})}_{\Delta\nu_{j,0}} -\\& \frac{1}{J_0} \sum_{j:D_j=0} \underbrace{(\expe{Y_{ij2}(0) - Y_{ij1}(0) | j(i) =j} - \expe{Y_{ij2}(0) - Y_{ij1}(0) | D_j})}_{\Delta\nu_{j,0}}.
\end{align*}

\noindent Letting $\Delta\nu_{j} = \Delta\nu_{j,0} + D_j \Delta\nu_{j,1}$ and $\Delta\epsilon_{ij} = \Delta\epsilon_{ij,0} + D_j \Delta\epsilon_{ij,1}$, it follows that $\widehat\tau$ takes the form (\ref{eqn: hatbeta-structural-model}) with $\beta = \tau_2$. 

\end{document}